\documentclass[aps,onecolumn,preprint,superscriptaddress,nofootinbib,floats]{revtex4-1}
\usepackage{amsmath,amssymb,color,mathrsfs, graphicx,verbatim,epsfig, bbm, wasysym}
\usepackage[hyperfootnotes=false]{hyperref}
\usepackage{slashed}
\allowdisplaybreaks

\def\lsim{\mathrel{\rlap{\lower4pt\hbox{\hskip1pt$\sim$}}
    \raise1pt\hbox{$<$}}}
\def\gsim{\mathrel{\rlap{\lower4pt\hbox{\hskip1pt$\sim$}}
    \raise1pt\hbox{$>$}}}

\newcommand{\be}{\begin{eqnarray}}
\newcommand{\ee}{\end{eqnarray}}

\def\addresses#1#2{\hbox to \hsize{\@tablebox{#1}\hfil\@tablebox{#2}}}
\def\@tablebox#1{\vtop{\hsize=5in \begin{flushleft} #1 \end{flushleft}}}

\def\beq{\begin{equation}}
\def\eeq{\end{equation}}
\def\bit{\begin{itemize}}
\def\eit{\end{itemize}}
\def\beqarray{\begin{eqnarray}}
\def\eeqarray{\end{eqnarray}}
\def\ttbar{$t\overline{t}$}
\def\ljets{$\ell$+jets}

\def\met{$\displaystyle{\not}E_T$}
\def\vecmet{$\vec{\displaystyle{\not}E}_T$}
\def\mtt{$m_{t\bar t}$}
\def\mtcl{$M_{T{\rm cl}}$}
\def\meff{$M_{\rm eff}$}
\def\PYTHIA{{\tt PYTHIA}}

\newcommand{\lp}{\big(\!}
\newcommand{\lb}{\big[\!}

\begin{document}

\baselineskip 0.6cm

\begin{titlepage}

\thispagestyle{empty}

\begin{flushright}
\end{flushright}

\begin{center}

\vskip 2cm

{\Large \bf Discriminating Top-Antitop Resonances using \\ \vspace{0.05in} Azimuthal Decay Correlations}

\vskip 1.0cm
{\large  Matthew Baumgart$^1$ and Brock Tweedie$^2$}
\vskip 0.4cm
{\it $^1$ Department of Physics and Astronomy, Johns Hopkins University, Baltimore, MD 21218} \\
{\it $^2$ Physics Department, Boston University, Boston, MA 02215} \\
\vskip 1.2cm

\end{center}

\noindent Top-antitop pairs produced in the decay of a new heavy
resonance will exhibit spin correlations that contain valuable
coupling information.  When the tops decay, these correlations imprint
themselves on the angular patterns of the final quarks and leptons.
While many approaches to the measurement of top spin correlations are
known, the most common ones require detailed kinematic
reconstructions and are insensitive to some important spin
interference effects.  In particular, spin-1
resonances with mostly-vector or mostly-axial couplings to top cannot
be easily discriminated from one another without appealing to
mass-suppressed effects or to more model-dependent interference with
continuum Standard Model production.  Here, we propose to probe the
structure of a resonance's couplings to tops by measuring the
azimuthal angles of the tops' decay products about the production
axis.  These angles exhibit modulations which are typically
$O$(0.1--1), and which by themselves allow for discrimination of
spin-0 from higher spins, measurement of the CP-phase for spin-0, and
measurement of the vector/axial composition for spins 1 and 2.
For relativistic tops, the azimuthal decay angles can be
well-approximated without detailed knowledge of the tops' velocities,
and appear to be robust against imperfect energy measurements and
neutrino reconstructions.  We illustrate this point in the highly
challenging dileptonic decay mode, which also exhibits the largest
modulations.  We comment on the relevance of these observables for
testing axigluon-like models that explain the top quark $A_{FB}$
anomaly at the Tevatron, through direct production at the LHC.

\end{titlepage}

\setcounter{page}{1}

\section{Introduction}

New resonances that decay to \ttbar\ are a common feature of many
models of physics beyond the Standard Model (SM).  Indeed, top's
unusually large coupling to the electroweak sector is often taken as a
hint of some special role in TeV-scale physics.

Recently, another hint has arisen at the Tevatron, in the measurement of
a 3.4$\sigma$ anomaly in the top pair forward-backward asymmetry
($A_{FB}$) at high invariant mass~\cite{Aaltonen:2011kc} (as well as lower-significance biases of the same sign
observed in~\cite{Aaltonen:2008hc,:2007qb,D0:2010afb}, and more recently in the dileptonic mode~\cite{CDF:2011afb}).  One possible
explanation of this effect is the presence of an $s$-channel
$q\bar{q}\to t\bar{t}$ resonance somewhere beyond the Tevatron's
direct discovery reach, which is interfering with the SM production.
As argued in~\cite{Frampton:2009rk}, a straightforward model framework
which can lead to the required interference is a color octet spin-1
boson with sizable axial couplings (of opposite sign) to light quarks
and tops; e.g., some variant of a heavy
axigluon~\cite{Pati:1975ze,Hall:1985wz,Frampton:1987dn}.

Such a resonance should be visible at the LHC, even if it turns out to
be very broad, and likely even in the upcoming 7 TeV
run~\cite{Bai:2011ed,Hewett:2011wz,Delaunay:2011gv,AguilarSaavedra:2011vw}.  This discovery would already be
extremely suggestive.  Direct confirmation of $A_{FB}$-related effects may also be possible,
despite the fact that the $pp$ initial state is forward-backward
symmetric~\cite{Antunano:2007da,Wang:2010du,Wang:2010tg,Xiao:2011kp,Hewett:2011wz}.
However, fully establishing the connection between the resonance and $A_{FB}$ will
benefit from independent measurements of the resonance's chiral couplings to top quarks
and light quarks, which will necessarily involve additional detailed
kinematic analyses both on- and
off-peak~\cite{Choudhury:2007ux,Godbole:2010kr,Degrande:2010kt,Cao:2010nw,Jung:2010yn}.
Most of these analyses exploit effects that, like $A_{FB}$ itself, arise from interference
with QCD.

Motivated by this endeavor, but also with an eye toward more general models, we can 
consider the following question:  Given that a resonance (either a peak or high-mass excess) is
discovered in the $m_{t\bar t}$ spectrum at the LHC, how much can we directly
learn about the chiral structure of its couplings to tops without relying
on more model-dependent QCD interference effects?  In particular,
in the case of a spin-1 resonance, could we determine if it couples to
tops mostly-axially by using only on-peak observables?

Of course the presence of a large axial component could be readily
established if the individual chiral couplings are highly biased
toward left-handed or right-handed.  At energies well above $2m_t$,
the different chiralities map almost uniquely onto distinct top
helicities, which can be discriminated by studying the rest frame
decay angles of individual tops~\cite{Frederix:2007gi,Bernreuther:2004jv}.  The differences
are also encoded in the lab-frame energies of the leptons and $b$-jets relative
to the top and to each other~\cite{Agashe:2006hk,Almeida:2008tp,Shelton:2008nq}, as well as 
in the energy sharing between the jets (or subjets) in hadronic decay~\cite{Krohn:2009wm}.

The situation becomes more subtle if the couplings to the two top
chiralities are more democratic.  Single-side spin analysis then
becomes inadequate to accurately probe the relative magnitudes between
vector and axial couplings, and we must incorporate measurements of
spin correlations between the two sides of the event.  These
correlations are affected by the interference between the different
helicity production modes, and can be used to directly probe the {\it
signed} ratio of top's chiral couplings.  Nonetheless, the most
commonly used top spin-correlation
variables~\cite{Barger:1988jj,Hara:1989yqa,Kane:1991bg,Mahlon:1995zn,Stelzer:1995gc,Mahlon:2010gw}
(reviewed in~\cite{Beneke:2000hk,Frederix:2007gi}) are largely
insensitive to helicity interference for heavy spin-1 resonances, and
generally involve complicated reconstructions that are not obviously
ideal for fast-moving tops.  Our main aim in this paper is to
demonstrate how to construct simple observables that {\it are}
sensitive, as well as straightforward to measure.

For spin-1 resonances well above $2m_t$, the on-peak helicity
interference is most directly encoded in the azimuthal distributions
of the tops' decay products about the \ttbar\ production axis.  This
is quite analogous to the effect measured at LEP in
$Z\to\tau^+\tau^-$~\cite{Bernabeu:1990na,Barate:1997mz,Abreu:1997vp,Volkert:1997ed}.
It is also closely related to the observables explored
in~\cite{Buckley:2007th,Buckley:2008pp,Buckley:2008eb,Murayama:2009jz,Boudjema:2009fz},
which use the same type of interference as a probe of the spins of new
pair-produced particles, or particles that are singly-produced in
association with a jet.  We now assume that we know the spin of the
top quark, and use the interference to tell us about how it was
produced.\footnote{See also~\cite{Kiyo:2000ag} for a study using the
individual top decay azimuthal angles to probe new \ttbar\ production
contributions at an ILC.}  As is typical for top spin observables, the
decay product most sensitive to the correlation is the charged lepton,
and we focus here on the dileptonic decay mode.  We also emphasize
that the same effects occur in the $\ell$+jets mode, though with
smaller amplitude.  We will show that the dileptonic azimuthal
distributions can be measured at the LHC, even given the usual
kinematic difficulties, and even using very minimalistic
reconstruction strategies.  These distributions exhibit surprisingly
robust modulations at the 30-40\% level for pure vector or axial
coupling, but with opposite signs.

Our observations also generalize beyond spin-1 resonances.  The minimal couplings to spin-2 have a similar chiral structure, which will be amenable to the same style of analysis.  Spin-0 resonances near \ttbar\ threshold have been a topic of detailed study for the past two decades, due to the possibility of detecting scalars from the Higgs-sector via gluon fusion (e.g.~\cite{Bernreuther:1997gs,Bernreuther:1998qv}).  Variables sensitive to the scalar/pseudo-scalar composition have therefore been known for a long time.  However, we will emphasize that azimuthal correlations make this composition visible in a very transparent way, especially if the resonance is heavy.  They also allow for a straightforward determination of whether a resonance is spin-0 versus higher spin, independent of \ttbar\ polar decay angle correlations or the production angle distribution.  Given the large amount of easily-measured information made available to us through these correlations, we highly advocate incorporating them into the set of \ttbar\ resonance discrimination observables laid out in~\cite{Frederix:2007gi}.

In the next section, we discuss the angular spin-correlation observables.  We start by reviewing the basic formalism, and then proceed to the details particular to spin-0, spin-1, and spin-2.  In particular, our discussions of spin-1 and spin-2 demonstrate how vector-coupled and axial-coupled resonances can be distinguished from each other with azimuthal correlations, as well as from spin-0.  In Section~\ref{sec:sim} we demonstrate that our observables should be accessible experimentally in the dileptonic channel.  We estimate the amount of cross section required to be able to distinguish between pure vector and axial-vector at early- and late-stage LHC, and discuss possible relevance to the Tevatron top $A_{FB}$ anomaly.  We conclude in Section~\ref{sec:conclusions}.  The Appendices include complete formulas for leading-order, on-peak \ttbar\ production and decay angular distributions for resonances up to spin-2.

\section{Spin Correlations}   
\label{sec:spinCorr}

After a heavy resonance decays to a pair of top quarks, the tops
themselves decay into $b\ell^+\nu$ or $bq\bar{q}'$ through an on-shell
$W$ boson.  In general, a given 6-body final-state kinematic
configuration could be produced by a number of intermediate
top/anti-top spin configurations.  Since we never measure the tops'
spins, these different spin channels coherently interfere with one
another, leading to distinctive imprints on the final-state angular
distributions.

The relevant behavior of the top decay matrix elements
becomes transparent when we exploit a fortuitous simplification.
If we look at the 3-body semileptonic top decay amplitude at
leading-order, stripping off the $W$ propagator, we can Fierz down to
\beq
\mathcal{M}(t\to b\ell^+\nu) \propto \Big( u(t)^T_L \, \epsilon \, v(\ell)_L \Big) \; 
\Big( u(b)^\dagger_L \, \epsilon \, u(\nu)^*_L \Big), \label{eq:Fierz}
\eeq
where $u$'s and $v$'s represent the spinor polarizations, and we have
kept only the two components from left-chirality.  Given a specific
top spin and decay configuration, we can map out how the amplitude
changes as we rotate the decay products in the top rest frame.
Because the $b$-quark and neutrino factorize off into a
rotationally-invariant product, the angular dependence of the
amplitude is determined exclusively by the direction of the
lepton:
\be
\mathcal{M}(t_\uparrow\to b\ell^+\nu) & \propto & e^{ i\phi_\ell/2} \cos\frac{\theta_\ell}{2} \nonumber \\
\mathcal{M}(t_\downarrow\to b\ell^+\nu) & \propto & e^{-i\phi_\ell/2} \sin\frac{\theta_\ell}{2},
\label{eq:topDecayME}
\ee
where we are using a standard right-handed coordinate system with spin
quantized along the $z$-axis.  If we instead decay an anti-top, we
simply swap $\cos \leftrightarrow \sin$.  (The same formulas also hold
with $\ell\to d/s$ in fully hadronic decay.)  Notably, the phases of
the amplitudes depend on the azimuthal angle of the lepton about the
top's spin vector.

Since we are ultimately interested in resonances beyond the Tevatron
reach, and well-above top threshold, we will restrict our
discussions to the chiral production limit.  (More complete formulas,
valid down to threshold, appear in the Appendices.)  In this limit, the
most natural spin basis is the helicity basis.  We therefore quantize
spin along the top/anti-top production axis in the \ttbar\ CM frame, and
measure $\phi$ around this axis  (the ``azilicity angle,'' in the terminology
of \cite{Gallicchio:2010dq}).  By convention we choose +$z$ to
point along the $t$ direction.  This may differ with conventions used in other
places, where the $\bar t$ decay coordinate system is independently defined with respect to its
own direction of flight.  To define the $x$- and $y$-axes, we
use the beams.  We can construct $\hat{y}$ by crossing the $t$
direction with one of the beam directions, so that it points out of 
the $t\bar t$ production plane.  (The ambiguity over which beam to pick at the
LHC will not be important for our purposes.)  The $\hat{x}$ direction is then defined
as usual for a right-handed coordinate system, as $\hat{y}\times\hat{z}$.  Consequently, when we
measure $\phi$ angles about the production axis, we define $\phi = 0$ to lie in the
production plane.  This is, of course, our
only physically-motivated choice.\footnote{In the presence of
nonvanishing $t\bar t$ system $p_T$, we must be somewhat careful when
performing this coordinate construction, as the beams are not
necessarily aligned in the \ttbar\ CM frame.  We address this issue in
Section~\ref{sec:sim}.}
Finally, we construct the individual top and anti-top decay
coordinate systems by actively boosting the tops to rest
(without rotation) along the production axis, and measuring all angles 
in this common reference frame.

Assembling a complete 6-body decay matrix element for a hypothetical
heavy particle (``$X$'') is now straightforward.  Taking the narrow
width approximation, and implicitly picking some definite initial spin
state, we can write
\beq
\mathcal{M}_{\rm tot}  =  \sum_{a,\bar a} \mathcal{M}(X \to t_a \bar t_{\bar a}) 
\mathcal{M}(t_a \to b\,\ell^+\nu) \mathcal{M}(\bar t_{\bar a} \to \bar b\,\ell^-\bar\nu),
\label{eq:mainMatrix}
\eeq
where $a$ and $\bar a$ label the spin of the top and anti-top,
respectively.  For spin-0 resonances, the tops are forced to have
opposite spins (identical helicities).  In the chiral limit, tops from
a heavy spin-1(2) decay will have identical spins (opposite
helicities).  The matrix elements schematically reduce to
\be
\mathcal{M}_{\rm tot}(X_0 \to b\,\ell^+\nu\bar b\,\ell^-\bar\nu) & \sim &
       \mathcal{M}(X_0\to t_\uparrow  \bar t_\downarrow) \, e^{ i(\phi_\ell - \overline{\phi}_\ell)/2}
                                                            \cos\frac{\theta_\ell}{2}\cos\frac{\bar\theta_\ell}{2} \, + \nonumber \\
  &  & \mathcal{M}(X_0\to t_\downarrow\bar t_\uparrow  ) \, e^{-i(\phi_\ell - \overline{\phi}_\ell)/2}
                                                           \sin\frac{\theta_\ell}{2}\sin\frac{\bar\theta_\ell}{2}   \nonumber \\
\mathcal{M}_{\rm tot}(X_{1(2)} \to b\,\ell^+\nu\bar b\,\ell^-\bar\nu) & \sim &  
       \mathcal{M}(X_{1(2)}\to t_\uparrow  \bar t_\uparrow  ) \, e^{ i(\phi_\ell + \overline{\phi}_\ell)/2}
                                                                 \cos\frac{\theta_\ell}{2}\sin\frac{\bar\theta_\ell}{2}  \, + \nonumber \\
  &  & \mathcal{M}(X_{1(2)}\to t_\downarrow\bar t_\downarrow) \, e^{-i(\phi_\ell + \overline{\phi}_\ell)/2}
                                                                 \sin\frac{\theta_\ell}{2}\cos\frac{\bar\theta_\ell}{2}, 
\label{eq:fullmatrix}
\ee
with the bar indicating the anti-top ($\ell^-$) decay angle.  The
final rate will therefore contain modulating terms dependent on
$\phi_\ell \pm \overline{\phi}_\ell$, with phase or magnitude
dependent on the complex production matrix elements.  For the case of
spin-0, which modulates as ($\phi_\ell-\overline{\phi}_\ell$), the two
$X_0$ decay elements are conjugates of each other.  Consequently, the modulation is
subject to a pure phase offset, where phases differing from 0
or $\pi$ signal CP-violation.  For spin-1 and spin-2, which modulate
as ($\phi_\ell+\overline{\phi}_\ell$), the effect is instead pure
magnitude, varying with the signed ratio of chiral couplings.  (E.g.,
in the case of purely LH or RH chiral production, there is only one
helicity channel, and therefore no modulation effect.)  
Since we will consider scenarios where $X$ is a singly-produced resonance, and not necessarily
extremely narrow, there may also be contributions to the azimuthal modulations from interference 
with the SM.  Still, the effect tends to be subleading for resonances with a well-defined peak.  
We discuss the issue in more detail at the end of Appendix~\ref{app:spin1}.

We can also see similar correlations between {\it any} two
decay products, one from the top side and one from the anti-top side.
While Eq.~\ref{eq:Fierz} naively suggests no dependence on the
$b$-quark and neutrino directions, these are still highly
kinematically correlated with the lepton via the (approximately) fixed
mass of the $W$ boson and overall four-momentum conservation for the
top.  Integrating out the phase space of all but one particle on each
side of the event leaves over angular correlations that are
essentially identical in structure to the double-lepton correlations,
including phase information.  The correlations simply scale with
constant spin analyzing powers associated to each particle species:
1 for leptons, -0.3 for neutrinos, -0.4 (+0.4) for 
$b$-quarks ($W$-bosons), and 0.5 for the softer of the two
$W$ decay products in the top rest frame~\cite{Brandenburg:2002xr}.
The effects which we seek to exploit are maximal for
leptons, but could also be measured in principle by including $b$-jets
or jets from the $W$ decay.  (Indeed, formulas analogous to
Eq.~\ref{eq:fullmatrix}, but instead using $W$-bosons, appeared
long ago in~\cite{Kane:1991bg} in the context of spin correlations
between top pairs at the SSC.)

Here, we focus on the double-lepton correlations, and
therefore the dileptonic $t\bar t$ decay channel.  This maximizes the
azimuthal modulation effects and is the least susceptible to reducible
backgrounds.  However, it also has the lowest rate (5\%, versus 30\%
for $\ell$+jets), and suffers from kinematic reconstruction difficulties
due to the two approximately back-to-back neutrinos.  We address the
latter difficulty in Section~\ref{sec:sim}.  Incorporating the
$\ell$+jets mode will also be very important, even though the modulations will be smaller
in amplitude (e.g., 40\% if we correlate a lepton with a $b$-quark or $W$-boson from
the other side of the event).  We reserve a detailed
exploration for future work, but include some commentary.

In what follows, we will usually be integrating out all but one or two angular variables, 
only restoring the full (and usually very complicated) correlated angular dependence in the Appendices.
Much of the discussion will parallel that of~\cite{Frederix:2007gi}, which comprehensively studied some of the standard
correlation observables in the context of resonance discrimination for arbitrary spins.
We also note that more sophisticated analyses, such as the likelihood approach of~\cite{Gao:2010qx,Melnikov:2011ai}, 
could potentially distinguish the different resonance scenarios even more efficiently.

\subsection{Spin-0}
\label{subsec:spin0}

We begin with the well-studied example of a spin-0 resonance.  Single-production
of such a resonance is typically chirality-suppressed from a $q\bar q$ initial state, and dominantly proceeds through
heavy quark loops in gluon fusion.  It is therefore perhaps nongeneric for this process to have a healthy rate 
above backgrounds when the resonance mass is at or above a TeV.  
However, independent of any model-building concerns, at a minimum, spin-0 serves as both an
illustrative example of our methods and as a baseline hypothesis against which any higher-spin
resonances must be tested.

At dimension-four, an arbitrary scalar coupling to $t \bar t$ is made of a linear
combination of CP even and odd pieces,
\beq
\mathcal{L}_{\rm int} = -\phi^a \, \bar{t}^i  \left( y + i\,\tilde{y} \, \gamma_5 \right) T^a_{ij} t^j,
\label{eq:spin0Lag}
\eeq
where $T^a_{ij}$ is the color matrix appropriate to the scalar's
representation (singlet or adjoint).  The dependence on the scalar's color
factorizes, so we ignore it in the remainder of the discussion.  
Since our interest is in the different top helicity
components, we can rewrite Eq.~\ref{eq:spin0Lag} in the chiral field basis,
\beq
\mathcal{L}_{\rm int} \to -y\, \phi \left(e^{i\, \alpha}\, \bar{t}_R t_L \,+\, e^{-i\, \alpha}\, \bar{t}_L t_R \right),
\label{eq:spin0LagAlt}
\eeq
where $y\,e^{i\, \alpha} \equiv y - i\, \tilde{y}$.
Since the $t \bar t$ pair can have no angular momentum along their
production axis, only same-helicity tops are created
in the scalar decay.  In the chiral limit, the production matrix elements differ only by the overall phase factor $e^{\pm i\,\alpha}$.
The angle $\alpha$ is equal to 0 for a CP-even scalar and $\pi/2$ for
a pseudoscalar.

Squaring $\mathcal{M}_{\rm tot}$ in Eq.~\ref{eq:mainMatrix} and integrating over phase space
except for lepton orientations,
\beq
\frac{d^4\Gamma}{d \Omega_\ell\, d \bar\Omega_\ell } \,\propto\, 
1 + \cos\theta_\ell \, \cos\bar\theta _\ell 
- \sin\theta_\ell \, \sin\bar\theta_\ell \, 
\cos \left( \phi_\ell - \bar\phi_\ell + 2\alpha \right).
\label{eq:niSpin0}
\eeq
Getting the distribution of polar angles is trivial from this equation,
as either lepton's azimuthal integral will kill the last term.  This leaves over
the well-known formula
\beq
\frac{d^2\Gamma}{d\cos\theta_\ell\, d\cos\bar\theta_\ell} \propto
1 + \cos\theta_\ell \, \cos\bar\theta_\ell.
\label{eq:dcSpin0} 
\eeq
The prefactor of the double-cosine term is characteristic of a scalar,
and has been advocated for distinguishing the overall spin of a
resonance~\cite{Frederix:2007gi} (though, we re-emphasize, typically with a coordinate change
$\bar\theta_\ell\to\pi-\bar\theta_\ell$).  However, we see that
dependence on the phase of the coupling has dropped out.  To retain
it, a traditional approach is to instead integrate out all angles
except for the 3D opening angle between the leptons, which we call $\chi$.\footnote{Elsewhere in the literature
this has been called $\phi$.  We relabel it to avoid confusion.}
Recall that we have actively boosted both tops
along the production axis into their respective rest frames, and then defined 
a common coordinate system with the $z$-axis given by the $t$ direction.  
It is in these coordinates where we measure $\chi$.
We get a simple linear dependence on $\cos\chi$,
\beq
\frac{d\Gamma}{d \cos\chi} \,\propto\,  1 + \frac{1-2\cos(2\alpha)}{3}\cos\chi.
\label{eq:chiEq}
\eeq
A CP-even scalar has a different slope compared to a pseudoscalar, $-\frac{1}{3}$ vs.~+1, 
making the distinction between the two straightforward in principle, provided that one can fully 
reconstruct the individual top quark rest frames.  Such a reconstruction can be challenging.  However,
as shown in~\cite{Bai:2008sk}, the resolution on the opening angle may still be good even for highly
boosted dileptonic top pairs.  
A more fundamental issue, as we will see in the next subsection, is that spin-1 and spin-2 resonances
all look identical to the CP-even scalar (and to each other) when using this angle.

A more straightforward way to maintain the $\alpha$-dependence, without needing polar angles,
is to integrate Eq.~\ref{eq:niSpin0} over all angles
except for the $\phi$ offset between the leptons.  This gives a modulating distribution
\beq
\frac{d\Gamma}{d(\phi_\ell-\bar\phi_\ell)} \,\propto\, 1 - \left( \frac{\pi}{4} \right)^2 
\cos (\phi_\ell - \bar\phi_\ell + 2\alpha), 
\label{eq:spin0}
\eeq
with a sizable amplitude of roughly 60\%.  We see that the modulation flips sign as we go from
CP-even scalar ($\alpha = 0$) to pseudoscalar ($\alpha = \pi/2$), and that the magnitude of the
modulation is the same $O$(1) size for any admixture.  Since
$\phi_\ell$ and $\bar\phi_\ell$ are boost-invariant along the individual tops'
directions of motion in the \ttbar\ CM frame, using Eq.~\ref{eq:spin0} may be amenable to
simpler $t\bar t$ reconstruction methods that do not necessarily need accurate information
about the tops' velocities.  Moreover, as we will see below, the equivalent distributions for
spin-1 and spin-2 resonances are flat, breaking the degeneracy with the CP-even scalar, and already
allowing a significant degree of spin discrimination in this variable alone.

\subsection{Spin-1}
\label{subsec:spin1}

We can write out the generic dimension-four coupling between a spin-1 resonance and a $t\bar{t}$ pair as
\beq
\mathcal{L}_{\rm int} = \, A^a_\mu \,  \bar{t}^i \left(g_V + g_A \gamma^5\right) \gamma^\mu T^a_{ij} t^j,
\label{eq:spin1Lag}
\eeq
where $g_V$ and $g_A$ are real numbers.
Again, we will suppress color in what follows, and go over to the chiral field basis,
\beq
\mathcal{L}_{\rm int} \to \, A_\mu \left(  g_L \, \bar{t}_L \gamma^\mu t_L + g_R \, \bar{t}_R \gamma^\mu t_R  \right).
\eeq
Unlike the couplings of a scalar, these couplings preserve chirality, and they differ in magnitude and sign
instead of being complex conjugates of each other.
For convenience, we define a coupling angle $\xi$,
\be
g_L &=& g \cos\xi \nonumber \\ 
g_R &=& g \sin\xi.
\ee
Our decay angular distributions will be functions of $\xi$, with the overall $g$ factored out.  By imposing 
symmetry under parity, we can have the special cases of $g_L = g_R$ (true vector, $\xi = \pi/4$)
or $g_L = -g_R$ (axial-vector, $\xi = 3\pi/4$).  We are particularly interested in finding observables that
discriminate between these two cases.

We again assume the chiral limit for top production, setting $m_t = 0$
in the decay matrix element of the heavy resonance.  
The finite-mass effects in the angular
distributions can in fact be different for the true vector and the
axial-vector, in principle providing additional means of
discrimination if the resonance is light enough.\footnote{In
particular, the double-cosine distribution for spin-1 resonances in
the \ttbar\ threshold limit is quite different between the two cases.
An axial-vector has a relative double-cosine coefficient of -1 for all
values of $m_t/M$, whereas a true vector's relative coefficient
approaches -1/3 at threshold ({\it cf.} Eq.~\ref{eq:finiteMtSpin1}).
In contrast, for a 1 TeV resonance, the coefficients differ from each
other by only about 10\%, and the discrepency shrinks like
$(m_t/M)^2$.  Similar conclusions also hold for the $\cos\chi$ distribution,
though the finite-mass effects there might be easier to measure since
the distribution is one-dimensional instead of two-dimensional.}  
However, by focusing on the features of the angular
correlations that are asymptotically insensitive to $m_t/M$, our
observations here will be robust up to arbitrarily high resonance
masses.

We begin our discussion of spin-1 angular distributions by looking at the standard double-cosine distribution, integrating
over all other angles,
\beq
\frac{d^2 \Gamma}{d\cos\theta_\ell \, d\cos\bar\theta_\ell} \propto
1 + \cos(2\xi) \left( \cos\bar\theta_\ell - \cos\theta_\ell \right)
- \cos\theta_\ell\,\cos\bar\theta_\ell.
\label{eq:spin1DoubleAngle}
\eeq
This is the direct analog of Eq.~\ref{eq:dcSpin0} in the spin-0 case.
The first term signals parity violation, and is only active if the
resonance is neither pure vector nor pure axial-vector.  
(Note that $\cos(2\xi) = (g_L^2-g_R^2)/(g_L^2+g_R^2)$.)
As discussed in
the introduction, a large chirality bias could be established already
with single-side angular distributions, such as
$d\Gamma/d\cos\theta_\ell$, and the construction of the double-cosine
distribution then essentially serves only as a cross check.
Incidentally, this observation by itself would already disfavor a
spin-0 interpretation.  On the other hand, in the more
chirality-symmetric case ($\cos(2\xi) \simeq 0$, $|g_L| \simeq
|g_R|$), the correlated distribution becomes more useful, as the sign
of the double-cosine term is opposite that for the scalar.  This is a
direct indication that the tops are coming out in a same-spin
(opposite-helicity) configuration, again disfavoring spin-0~\cite{Frederix:2007gi}.
Crucially, however, all terms in the distribution are insensitive to
the relative sign between the chiral couplings, up to the finite-mass
corrections mentioned above.  Consequently, it will likely be quite
difficult to tell whether the resonance is vector-coupled or
axially-coupled using only the double-cosine distribution.

Turning to the distribution over the 3D lepton angle $\chi$ does not help.  It is
\beq
\frac{d\Gamma}{d \cos\chi} \,\propto\,  1 - \frac{1}{3}\cos\chi,
\label{eq:spin1Chi}
\eeq
independent of $\xi$.  This is identical to the CP-even scalar distribution, Eq.~\ref{eq:chiEq}, again up
to the small $(m_t/M)^2$ corrections which we have neglected.
We therefore learn very little from this distribution if we are interested in the properties of spin-1 resonances.

To robustly determine the signed ratio of chiral couplings, we again
suggest the alternative tactic of focusing on the lepton azimuthal
angles with respect to the production axis.  There are now three major
differences with the spin-0 case.  First, as indicated in
Eq.~\ref{eq:fullmatrix}, the modulation switches from $(\phi_\ell -
\bar\phi_\ell)$ to $(\phi_\ell + \bar\phi_\ell)$, and therefore now
refers to the orientation of the production plane
($zx$-plane).\footnote{One way to picture the construction of
$(\phi_\ell + \bar\phi_\ell)$ is as follows.  Reflect the $\ell^-$ in
the \ttbar\ production plane, thereby flipping the sign of its $\phi$ angle.  Then
take the difference in $\phi$ as before.}  This is a consequence of
the fact that the spin-1 resonance carries a polarization which is
necessarily fully correlated with the beams.  The second difference is
that the coupling angle ($\alpha\to\xi$) now affects the modulation's
amplitude, not its phase.  The third difference, which also follows
from nonzero polarization, is that the decay angle distributions can
become correlated with the \ttbar\ production angle, which we call $\Theta$.  
In what follows, we will implicitly average between rates at $\Theta$ and $\pi-\Theta$,
so that effectively $\Theta \in [0,\pi/2]$.

For a resonance produced in $q\bar q$ annihilation, the component of
the resonance's spin along the beam is $J_{\rm beam} = \pm 1$, and the
joint production/azimuthal angle distribution is
\beq
\frac{d^2\Gamma_{J_{\rm beam}=\pm1}}{d(\phi_\ell+\bar\phi_\ell)\, d \cos\Theta} 
\,\propto\,  \left(1+\cos ^2\Theta\right) - \left(\frac{\pi}{4}\right)^2 \sin(2\xi) \sin^2\Theta \cos(\phi_\ell+\bar\phi_\ell).
\label{eq:polarIntSpin1}
\eeq
Unlike the double-cosine distribution, which depends only on $\cos(2\xi)$, the azimuthal modulation
depends on $\sin(2\xi) = 2g_Lg_R/(g_L^2+g_R^2) = 2\,/(g_L/g_R+g_R/g_L)$, and
hence is directly sensitive to the signed ratio $g_L/g_R$.
Perhaps fortunately, the modulation is largest for central production, which
is the easiest case to reconstruct in real detectors.  (The modulation
shuts off for \ttbar\ produced along the beam, since only one helicity
state can contribute.)  Fully
integrating out the production angle, the distribution reduces to
\beq
\frac{d\Gamma_{J_{\rm beam}=\pm1}}{d(\phi_\ell+\bar\phi_\ell)} 
\,\propto\,  1 - \frac12\left(\frac{\pi}{4}\right)^2 \sin(2\xi)  \cos(\phi_\ell+\bar\phi_\ell),
\label{eq:spin1}
\eeq
which is very similar to the spin-0 $\phi$-{\it difference}
distribution, Eq.~\ref{eq:spin0}.  However, it is straightforward to
see that, respectively, a spin-1 resonance is non-modulating over
$(\phi_\ell - \bar\phi_\ell)$,\footnote{This modulation is generically
induced by nonzero $m_t$ effects ({\it cf.} Eq.~\ref{eq:appAzCorrOnly0}).}  
whereas a spin-0 resonance is
non-modulating over $(\phi_\ell + \bar\phi_\ell)$.  Using {\it both}
of these distributions can therefore tell us quite a lot about both
the spin of the resonance and its coupling structure, independent of
any other measurements such as the production angle distribution
(trivial for spin-0 but nontrivial for spin-1), or observables
incorporating the tops' polar decay angles (sensitive to chirality
biases).

We can also give the analogue of Eq.~\ref{eq:polarIntSpin1} for a
resonance with $J_{\rm beam}=0$.  Such a state would come from gluon
fusion and is typically suppressed, partly because 
the gluon parton luminosities drop very quickly with energy, and partly because
the required gauge-invariant operators are dimension-six.\footnote{Note that the Landau-Yang theorem \cite{Landau:1948ab,Yang:1950rg} forbids
production of a color-singlet spin-1 resonance through gluon fusion.  Implicitly,
we only consider color octet resonances.}  We include this case
for completeness, and since it is still possible that the light quark couplings
are highly suppressed, making this the dominant production mode.
The production/azimuthal angle distribution becomes
\beq
\frac{d^2\Gamma_{J_{\rm beam}=0}}{d(\phi_\ell+\bar\phi_\ell)\, d \cos\Theta} \,\propto\,
\sin ^2\Theta \left[1+\left(\frac{\pi}{4}\right)^2 \sin(2\xi)  \cos(\phi_\ell+\bar\phi_\ell) \right].
\label{eq:polarIntSpin1Pol0}
\eeq
We see that the modulation persists, but it is now factorized from the production angle and is consequently
twice as large when $\Theta$ is integrated out.  It also has the opposite sign.
While this change in sign and magnitude could certainly lead to confusion over the true value of $g_L/g_R$,
the production angle distribution is nonetheless still quite distinct, allowing an experimentally-driven
determination of whether a $J_{\rm beam}=0$ or a $J_{\rm beam}=\pm 1$ interpretation is appropriate.

\subsection{Spin-2}
\label{subsec:spin2}

Like spin-1, spin-2 particles can couple with different magnitudes to left and right-handed tops (and other fermions), though the leading interaction
is now dimension-five,
\beq
\mathcal{L} \,\propto\,  \frac{\cos\xi}{\Lambda}h^{\mu\nu} \left( \bar t_L (i \overleftrightarrow{D}_\nu ) \gamma_\mu t_L  \right) +
 \frac{\sin\xi}{\Lambda}h^{\mu\nu} \left( \bar t_R (i \overleftrightarrow{D}_\nu ) \gamma_\mu t_R  \right), 
\label{eq:spin2Lag}
\eeq
where $h$ is a traceless and symmetric rank-2 tensor field, and $\Lambda$ is the
heavy scale that controls the strength of the operator.  (This coupling
is for a color-singlet field, but for color-octet we need only a trivial
color matrix insertion.)  Once again, we can build the traditional
observables and see that they tell us nothing about the
relative sign between left and right-handed couplings.
(The double-cosine distributions and $\cos\chi$ distributions are identical
to spin-1.)  Just as with
the other spins, the azimuthal angle distributions preserve this information.  

Quite analogous to spin-1, the top production and decay angle distributions are correlated with each other.  To
get the equivalent of Eq.~\ref{eq:polarIntSpin1}, we must replace
\beqarray
1 + \cos^2\Theta & \to & 1 - 3\cos^2\Theta + 4\cos^4\Theta \nonumber \\
    \sin^2\Theta & \to & -\left( 1 - 5\cos^2\Theta + 4\cos^4\Theta \right).
\eeqarray
The coefficient of the modulating term now goes through a zero at $\cos\Theta=0.5$.  Integrating
over production angles, the modulation is diluted with respect to spin-1 by a factor of three (and flipped in sign),
\beq
\frac{d\Gamma_{J_{\rm beam}=\pm1}}{d(\phi_\ell+\bar\phi_\ell)} 
\,\propto\,  1 + \frac16\left(\frac{\pi}{4}\right)^2 \sin(2\xi)  \cos(\phi_\ell+\bar\phi_\ell).
\label{eq:spin2qqbar}
\eeq
While the production angle distribution is potentially a giveaway that the resonance is spin-2, 
the reduced $(\phi_\ell+\bar\phi_\ell)$ modulation will make discrimination of the top couplings much more difficult.
However, given adequate statistics, focusing on the region of central production would allow a much larger
modulation to be seen.

A spin-2 resonance can also be produced in gluon fusion, also through dimension-five operators.
This can in principle lead to both $J_{\rm beam}=\pm2$ and $J_{\rm beam}=0$, but the latter production
mode is in fact absent at dimension-five.\footnote{For color singlet spin-2, the
interaction $h_{\mu \nu} {\rm Tr} ( G^{\mu \alpha} G^\nu_\alpha )$, 
where $G$ is the gluon field-strength, contains no coupling
to the $J_{\rm beam} = 0$ state on-shell.  We could instead try to couple through the parity-odd interaction $h_{\mu \nu} {\rm Tr} ( G^{\mu
\alpha} \tilde{G}^\nu_\alpha )$, but this vanishes identically when $h_{\mu \nu}$ is traceless.  Considering an octet resonance, our only
options are
${\rm Tr} ( h_{\mu \nu} \left\{ G^{\mu \alpha}, G^\nu_\alpha \right\}
)$ or ${\rm Tr} ( h_{\mu \nu} [ G^{\mu \alpha}, \tilde{G}^\nu_\alpha ]
)$, each of which also lead to vanishing amplitude for $J_{\rm beam} = 0$.}
We therefore exclusively focus on $J_{\rm beam}=\pm2$.
(We include $J_{\rm beam}=0$ in our discussions in Appendix~\ref{app:spin2},
as it may still arise from yet higher-dimension operators.)  The production angle no longer
factorizes out, and we are left with a joint production/azimuthal angle distribution much
like Eq.~\ref{eq:polarIntSpin1}, but with an additional overall factor of $\sin^2\Theta$.
Since the production is biased more centrally, where the modulation is largest,
the $\Theta$-integrated modulation is now {\it enhanced},
\beq
\frac{d\Gamma_{J_{\rm beam}=\pm2}}{d(\phi_\ell+\bar\phi_\ell)} 
\,\propto\,  1 - \frac23\left(\frac{\pi}{4}\right)^2 \sin(2\xi)  \cos(\phi_\ell+\bar\phi_\ell).
\label{eq:spin2gg}
\eeq
Similar to the spin-1 case, gluon fusion leads to larger modulations with opposite sign.
Again, distinguishing between $q\bar q$ and $gg$ (as well as spin-1 in any production mode) 
is possible in principle by studying the $\Theta$ distribution.

\section{Observing Azimuthal Correlations at the LHC}  
\label{sec:sim}

We now turn to the question of whether the azimuthal modulations discussed
in Section~\ref{sec:spinCorr} are actually observable at the LHC, and 
in what cases we might be able to discriminate between different resonances.
We focus on the dileptonic mode, which has the largest azimuthal correlation effects
amongst \ttbar\ decay modes, but which also has the lowest branching
fraction and, more pressingly, the fewest number of visible decay
products.  Given the presence of {\it two} experimentally-invisible
neutrinos, it is fair to ask 1)~whether the resonance can be
identified, and 2)~whether the azimuthal observables discussed
above can be reliably reconstructed.  We address these two concerns in
turn, in the context of simple LHC simulations that include
showering/hadronization, gaussian energy smearing, and basic analysis
cuts.  As described in more detail in the subsections below, we find
that minimalistic strategies work quite well, and that the modulations are very
robust. \enlargethispage{-20mm}

\subsection{Simulation and basic event selection in the dileptonic mode}

We generate 6-body dileptonic resonance and background samples at leading-order, with full angular correlations, using {\tt MadGraph/MadEvent v4.4.51}~\cite{Alwall:2007st} and its topBSM~\cite{Frederix:2007gi} model.\footnote{We do not include interference effects between the resonance and Standard Model in the ensuing analysis, since we expect them to be largely washed-out when integrating over the peak and taking the absolute value $|\phi_\ell + \bar\phi_\ell|$.  Nonetheless, we have explicitly checked a small set of realistic, fully interfered samples with 15\% resonance width, pure vector couplings to light quarks (to maximize the persistence of the interference given the chirality-averaged initial state), and vector or axial couplings to tops.  We find that with $S/B \simeq 1$ in the resonance region, our azimuthal distributions are consistent with an incoherent average of the signal and background.}  We subsequently shower/hadronize with {\tt PYTHIA 6.4}~\cite{Sjostrand:2006za} and cluster the resulting particles into jets with {\tt FastJet v2.4.2}~\cite{Cacciari:2005hq} without detector simulation.  We then apply gaussian energy smearings $\sigma(E)/E$ to the jets ($5.6{\rm \, GeV}/E \oplus 1.25\sqrt{{\rm GeV}/E} \oplus 0.033$), electrons ($0.02$), and muons ($0.05\times\sqrt{E/{\rm TeV}}$).

For very energetic tops, the lepton may be separated from the $b$-jet by a fairly small angle, possibly failing ordinary isolation criteria and subjecting us to semileptonic heavy flavor backgrounds.  Nonetheless, these leptons can probably be reliably discriminated using tracker-level ``mini-isolation'' criteria with shrinking cones~\cite{Rehermann:2010vq,ATLAS:2010boo}, or through a number of other simple means~\cite{Thaler:2008ju,Brooijmans:2009boo}.  We accept into our analysis events with exactly two opposite-signed leptons (electrons or muons) that pass mini-isolation as described in~\cite{Rehermann:2010vq}, and with $p_T > 20$ GeV and $|\eta| < 2.5$.  This set of cuts (or some modest modification thereof) should be adequate to reduce heavy flavor backgrounds to a very subdominant level, while keeping the majority of the signal.

We reconstruct jets using the anti-$k_T$ algorithm~\cite{Cacciari:2008gp} with $R = 0.4$, keeping those with $p_T > 50$ GeV and $|\eta| < 2.5$.  No $b$-tags are applied.  Each lepton is then paired with the hardest jet which satisfies $\Delta\phi(j,\ell) < \pi/2$ and $m(j,\ell) < 200$ GeV.\footnote{For genuine boosted top decays, the $b$-jets are quite narrow, and their directions should be easy to accurately measure using the tracker.  We therefore assign no error beyond energy smearing in the reconstruction of the jet-lepton invariant mass, nor in any of our subsequent kinematic reconstructions.}  Cases with too few or non-unique pairings are discarded.

Our missing energy reconstruction is very naive, and hopefully conservative:  we simply balance the summed $\vec{p}_T$ of the two leptons and the two ``$b$-jets.''

We find that reducible backgrounds from $\ell^+\ell^-$+jets (including taus) and $W^+W^-$+jets are quite small throughout our analysis after all cuts, each at the 1--10\% level with respect to the irreducible dileptonic \ttbar.\footnote{To reduce the former background to this level, we require a simple $Z$-veto that rejects same-flavor lepton pairs with $m(\ell^+,\ell^-) = [75,105]$ GeV.  This has a very minor effect on our signal, but order-of-magnitude effect on $\ell^+\ell^-$+jets.}  In what follows, we explicitly keep only the irreducible background, which is by far our biggest concern. \enlargethispage{-30mm}

\subsection{Reconstruction of \mtt}

The most efficient way to separate our signal from the continuum \ttbar\ background is to measure \mtt\ and isolate the resonance region.  However, since we will investigate spin correlations in the dileptonic channel, we face an inevitable complication due to the two missing neutrinos.  Of course, by the time a dileptonic analysis becomes sensitive to the resonance, we should already know the resonance's mass (and perhaps also its width) quite well from measurements in $l$+jets and all-hadronic modes.  So if we can reconstruct \mtt, even approximately, then we already know where to look.

The two neutrinos represent six lost kinematic degrees of freedom, only two of which are recovered through the measurement of \vecmet.  Complete kinematic reconstruction is nonetheless possible by demanding on-shell tops and $W$s on the two sides of the event, which fix the other remaining four degrees of freedom through four constraints.  For energetic top pairs, the possible kinematic ambiguity regarding which $b$-jet belongs to which lepton is not a major issue, and solving for the complete system essentially reduces to finding the zeros of a quartic polynomial~\cite{Sonnenschein:2006ud}.  There are always four complex solutions, of which either zero, two, or four might be real.\footnote{A more realistic experimental analysis might instead solve the system using a multivariate likelihood approach (see, e.g.,~\cite{Affolder:2000vy}), allowing all of the measured transverse observables (\vecmet, $\vec{p}_T(l_i)$, $\vec{p}_T(j_i)$) to vary, approximately within their errors, along with the unknown neutrino 3-momenta.  This always gives real solutions by construction, and multiple solutions can be ranked or weighted by their likelihoods.}

This type of direct solution approach is not without its dangers.  The continuum \ttbar\ background at the LHC rises steeply toward small \mtt, and even a modest rate of misreconstruction can ``avalanche'' low-mass backgrounds into our high-mass signal region.  Indeed, the full kinematic solution can be highly sensitive to the precise values of visible kinematic variables, as well as to the detailed choice of top and $W$ masses used to find the solutions.\footnote{Even the natural Breit-Wigner scatter in the true event-by-event top and $W$ masses produces sizable tails in the reconstructed \mtt\ at parton-level.}  The situation is particularly difficult for energetic tops, as all of the decay products, including the neutrinos, approximately fall on a single axis in the transverse plane, and the neutrinos are roughly back-to-back.  Accurate event reconstruction then demands exceptionally good measurement of missing energy components.

The alternative is to consider reconstructions which are safer from the perspective of overestimating the mass of backgrounds, but which do so by actively attempting to {\it underestimate} the true mass.  A well-known example (advocated for dileptonic \ttbar\ resonance reconstruction in~\cite{Baur:2007ck}) is the cluster transverse mass,
\beq
M_{T{\rm cl}}^2 \equiv \left( \sqrt{p_{T{\rm vis}}^2 + m_{\rm vis}^2} + \displaystyle{\not}E_T \right)^2  -  \left(\vec p_{T{\rm vis}} + \vec{\displaystyle{\not}E}_T  \right)^2,
\eeq
with the ``vis'' subscript referring to the 4-vector sum of the two leptons and two $b$-jets.  This represents the smallest \mtt\ consistent with the measured \vecmet, without any on-shell criteria for the tops and $W$s.\footnote{More specifically, this is realized by a kinematic configuration where the neutrinos are exactly collinear in 3D (the precise energy-sharing is irrelevant), are aligned with \vecmet\ in the transverse plane, and have rapidity matching that of the visible cluster.  For generalizations of this method to cases where the missing particles are massive (they should again be set 3D-collinear, but all moving with the same velocity), see~\cite{Konar:2008ei,Konar:2010ma}.}  We can also try to improve upon this a bit by utilizing the fact that the neutrinos should be roughly collinear with the visible products when the tops are boosted.  To this end, we construct a kind of minimal single neutrino, which under-reconstructs the complete $\nu\bar\nu$ subsystem, by setting $\vec{p}_T(\nu) \equiv \vec{\displaystyle{\not}E}_T$ and setting $\eta$ to match that of the closest lepton in $\phi$.  (Setting $\eta$ to match the rapidity of the closest $b+l$ system yields nearly equivalent results.)  In addition, we could consider the \meff\ variable used in supersymmetry searches, which simply adds up the scalar $p_T$s of all visible products and \met.  \enlargethispage{-20mm} We could also choose to be even more minimalistic, and just take the invariant mass of the visible component of the event.\footnote{Perhaps the {\it most} minimalistic strategy, simply applying a hard cut to the $p_T$ of the leptons, was used in~\cite{Davoudiasl:2009jk}. This is fairly inefficient, so we do not pursue it here.  However, in cases where the resonance has enough cross section, this may be a clean and effective way to get a reliable signal sample.  We note that for highly boosted tops, the lepton $p_T$ is largely uncorrelated with the azimuthal angles about the production axis, and we suspect that our correlation measurements will continue to work even with such a manifestly kinematically-biased cut.}

To determine a reasonable method for our present work, amongst the many available options, we study a handful of kinematic reconstructions.  In addition to the four undercomplete reconstructions just described, we also consider two methods based the complete reconstruction using mass constraints, for which we must solve a quartic equation.  While the quartic will practically never yield exactly one real solution, and often yields zero real solutions, we find that a simple and conservative workaround is to take the real components of the neutrino momenta returned by the solutions, and pick the configuration that yields the smallest \mtt.  We also include the strategy of Bai and Han~\cite{Bai:2008sk}, which discards solutions that give the neutrinos too large a fraction of the scalar $p_T$ of the event, or which have imaginary components that are too large.  Complex solutions are then made real by taking the complex norm of the reconstructed top and anti-top 4-momentum components, and all acceptable solutions are included with equal weight.  (Complex solutions are only counted once.)  Equivalently for our purposes, we just pick one solution at random.

\begin{figure}[tp]
\begin{center}
\epsfig{figure=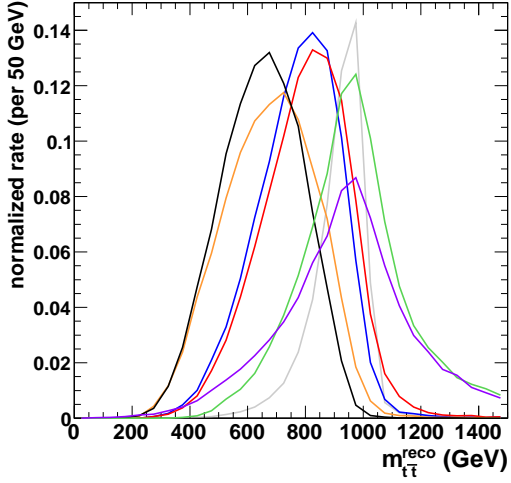,scale=0.40} \hspace{0.0cm}
\epsfig{figure=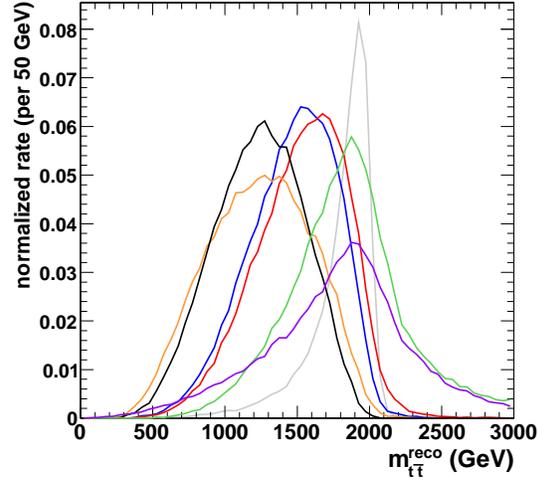,scale=0.40}
\caption{\it  Various approximations to \mtt\ as applied to narrow 1 and 2 TeV vector resonances.  Neutrino reconstructions include perfect (grey, scaled vertically by 1/2), \mtcl\ (blue), minimal $\eta$-collinear (red), visible-only (black), \meff\ (orange), minimal real-part quartic (green), and Bai-Han quartic (purple).}
\label{fig:MrecoSignal}
\end{center}
\end{figure}
\begin{figure}[tp]
\begin{center}
\epsfig{figure=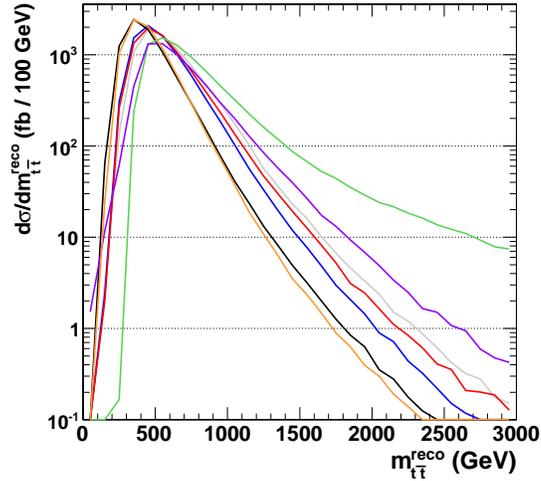,scale=0.40}
\caption{\it  Various approximations to \mtt\ as applied to continuum Standard Model \ttbar\ at LHC14.  Neutrino reconstructions include perfect (grey), \mtcl\ (blue), minimal $\eta$-collinear (red), visible-only (black), \meff\ (orange), minimal real-part quartic (green), and Bai-Han quartic (purple).}
\label{fig:MrecoBG}
\end{center}
\end{figure}

\begin{figure}[tp]
\begin{center}
\epsfig{figure=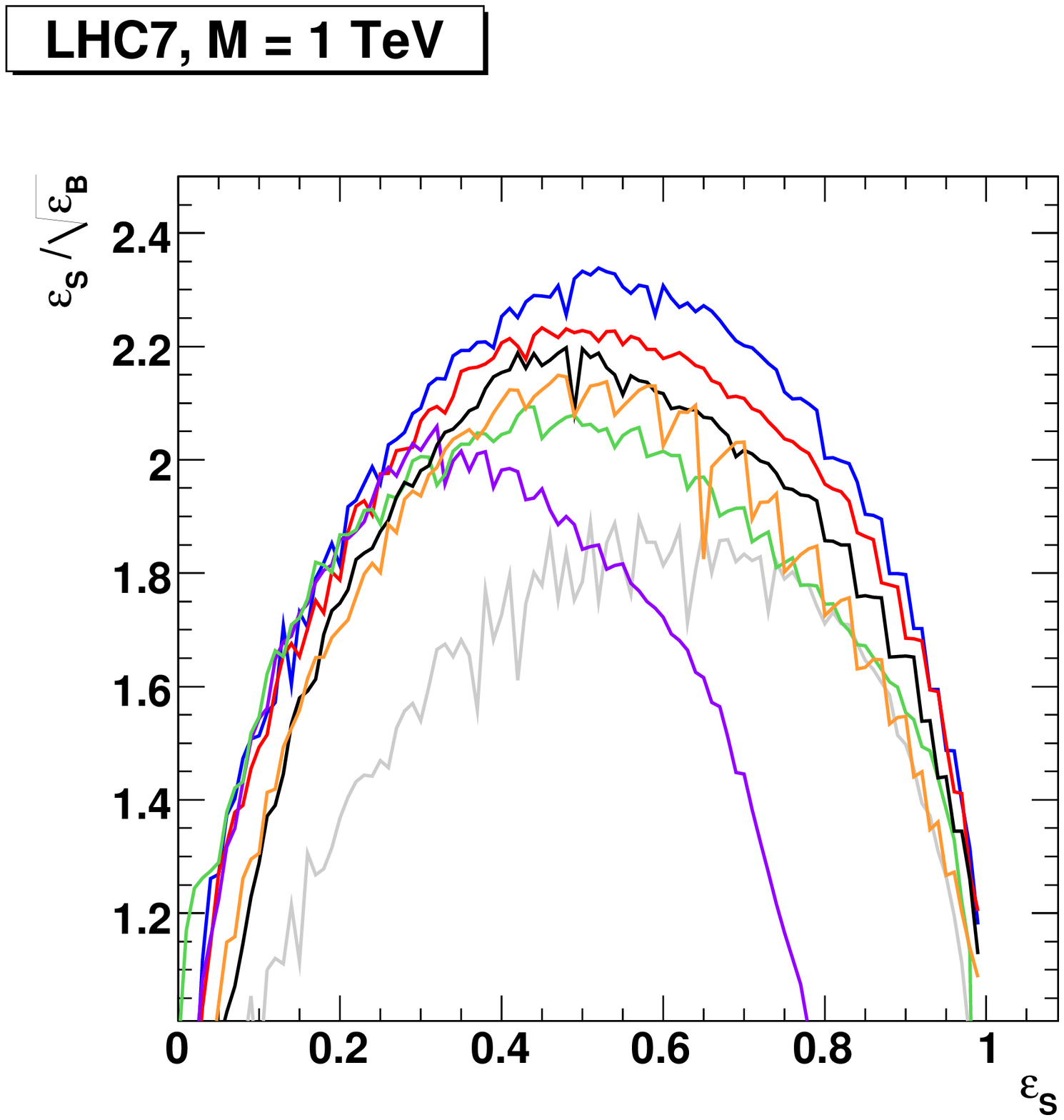,scale=0.40} \hspace{0.0cm}
\epsfig{figure=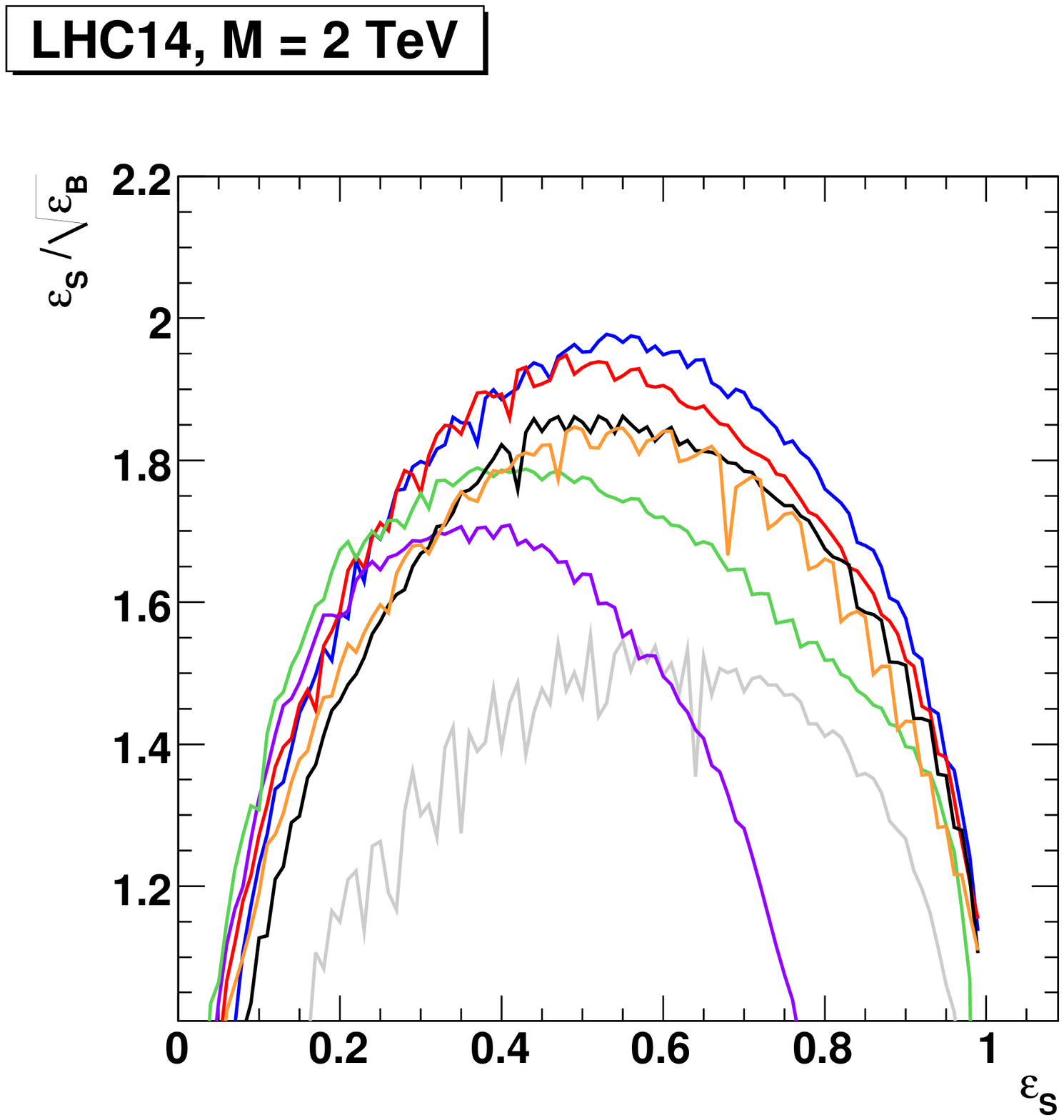,scale=0.40}
\caption{\it  Significance improvement versus signal efficiency, using optimized mass windows and transversity cuts, for a 1 TeV narrow resonance at LHC7, and for a 2 TeV narrow resonance at LHC14.  Neutrino reconstructions include perfect (grey, scaled to (1/2)$\varepsilon_S/\sqrt{\varepsilon_B}$), \mtcl\ (blue), minimal $\eta$-collinear (red), visible-only (black), \meff\ (orange), minimal real-part quartic (green), and Bai-Han quartic (purple).  (The 2 TeV curves were generated using a background sample with $m_{t\bar t} > $ 800 GeV at parton-level, to improve statistics.  Therefore the regions near 100\% efficiency are not fully accurate.  They would have a much steeper slope in reality, and the overall SIC levels on the rest of the plot would be higher.  However, even with modest cuts, sensitivity to this generator-level cutoff largely disappears.)}
\label{fig:SIC}
\end{center}
\end{figure}

In Figs.~\ref{fig:MrecoSignal} and~\ref{fig:MrecoBG}, we show the results of our reconstructions on narrow 1 and 2 TeV spin-1 vector-coupled resonances and on the SM continuum at the 14~TeV LHC, including an additional reference reconstruction where the neutrino 3-vectors are perfectly measured.\footnote{Axial resonances look nearly identical, while purely RH chiral resonances tend to have slightly harder distributions, and purely LH slightly softer.}$^,$\footnote{The reconstruction with perfectly measured neutrinos displays a pronounced low-mass tail, partially due to additional neutrinos lost in semileptonic $B$-hadron decays, and partially due to FSR of gluons off of the tops.  The shape can be improved with the inclusion of extra jets into the reconstruction, though we have erred on the conservative side and not done so.}  \enlargethispage{-10mm} Roughly speaking, there is a tradeoff between how tightly we attempt to sculpt the core of the signal peak and how often we overshoot the true mass.  To assess what this means in terms of discriminating signal from background, we run a simple cut optimization study, scanning over possible mass windows.  Inspired in part by the observations of~\cite{Bai:2008sk}, we also scan over a cut on the event's ``transversity,'' defined as the ratio of either \mtcl, \meff, or min($\{p_T(b,l)\}$) to the reconstructed mass.  In particular, this type of cut aids in removing a substantial contribution from gluon fusion backgrounds, which tend to peak at low production angles.  It also significantly cuts down the unphysical high-mass broadening of the background for the quartic reconstructions.

Some representative results for the narrow resonance optimizations are presented in Fig.~\ref{fig:SIC}, using the SIC (``significance improvement characteristic,'' i.e.\ $\varepsilon_S/\sqrt{\varepsilon_B}$) visualization of \cite{Gallicchio:2010dq}.  In practice, we find that \meff\ usually works best for the transversity cut, and we use it for all of our displayed SIC curves except for that of \meff\ itself, for which we use min($\{p_T(b,l)\}$).  We find that the consistently most powerful mass estimator is \mtcl, in combination with the cut on \meff/\mtcl.  We therefore use these to define the signal regions for our subsequent azimuthal angle analysis, maximizing $\varepsilon_S/\sqrt{\varepsilon_B}$.  Tables~\ref{tab:eff7} and~\ref{tab:eff14} contain the results of the optimization.

\begin{table}
 \centering
\begin{tabular}{|l|c|c|c|c|c|}  \hline
 LHC$\,$7   &  Mass/Transversity Cuts  &  $\varepsilon_S^{\rm selection}$ &  $\varepsilon_S^{\rm selection+cuts}$  &  $\sigma_B$ \\ \hline
 \ 1 TeV, Narrow     \ \ & \ $M_{T{\rm cl}} = [ 750,1025]$ GeV, $M_{\rm eff}/M_{T{\rm cl}} > 0.65$ \ \ & \ 0.38 \ \ & \ 0.24 \ \  &  \   21 fb \ \ \\ 
 \ 1 TeV, 15\%-Width \ \ & \ $M_{T{\rm cl}} = [ 700,1450]$ GeV, $M_{\rm eff}/M_{T{\rm cl}} > 0.65$ \ \ & \ 0.33 \ \ & \ 0.19 \ \  &  \   24 fb \ \ \\ \hline
 \ 2 TeV, Narrow     \ \ & \ $M_{T{\rm cl}} = [1600,2100]$ GeV, $M_{\rm eff}/M_{T{\rm cl}} > 0.50$ \ \ & \ 0.58 \ \ & \ 0.20 \ \  &  \ 0.07 fb \ \ \\
 \ 2 TeV, 15\%-Width \ \ & \ $M_{T{\rm cl}} = [1425,2925]$ GeV, $M_{\rm eff}/M_{T{\rm cl}} > 0.75$ \ \ & \ 0.40 \ \ & \ 0.09 \ \  &  \ 0.17 fb \ \ \\
\hline
\end{tabular}
\caption{\it Optimized mass window and transversity cuts, and the corresponding signal efficiencies and background cross sections, for the 7 TeV LHC.  Signal efficiencies are for a vector-like spin-1 resonance in the dileptonic mode, after basic selection and then after application of the cuts.}
\label{tab:eff7}
\end{table}
\begin{table}
 \centering
\begin{tabular}{|l|c|c|c|c|c|}  \hline
 LHC14   &  Mass/Transversity Cuts  &  $\varepsilon_S^{\rm selection}$ &  $\varepsilon_S^{\rm selection+cuts}$  &  $\sigma_B$ \\ \hline
 \ 1 TeV, Narrow     \ \ & \ $M_{T{\rm cl}} = [ 725,1000]$ GeV, $M_{\rm eff}/M_{T{\rm cl}} > 0.75$ \ \ & \ 0.35 \ \ &  \ 0.19 \ \ &  \ 179 fb \\ 
 \ 1 TeV, 15\%-Width \ \ & \ $M_{T{\rm cl}} = [ 700,1300]$ GeV, $M_{\rm eff}/M_{T{\rm cl}} > 0.75$ \ \ & \ 0.32 \ \ &  \ 0.16 \ \ &  \ 253 fb \\ \hline
 \ 2 TeV, Narrow     \ \ & \ $M_{T{\rm cl}} = [1475,2050]$ GeV, $M_{\rm eff}/M_{T{\rm cl}} > 0.60$ \ \ & \ 0.56 \ \ &  \ 0.27 \ \ &  \ 5.6 fb \\
 \ 2 TeV, 15\%-Width \ \ & \ $M_{T{\rm cl}} = [1425,2425]$ GeV, $M_{\rm eff}/M_{T{\rm cl}} > 0.60$ \ \ & \ 0.48 \ \ &  \ 0.19 \ \ &  \ 7.5 fb \\
\hline
\end{tabular}
\caption{\it \it Optimized mass window and transversity cuts, and the corresponding signal efficiencies and background cross sections, for the 14 TeV LHC.  Signal efficiencies are for a vector-like spin-1 resonance in the dileptonic mode, after basic selection and then after application of the cuts.}
\label{tab:eff14}
\end{table}

Of course, any conclusions drawn from such a simple study can only be taken as suggestive, but it does indicate that attempts to make detailed reconstructions of the neutrinos might be counterproductive, unless cleverer techniques are brought to bear.\footnote{We emphasize that the underperformance of the full quartic solutions with respect to \mtcl\ in this study is not just an artifact of our somewhat pessimistic missing energy reconstruction.   The conclusion remains unchanged if we switch to perfect \vecmet\ measurement and turn off energy smearings.}  It is also not clear that the maximum $\varepsilon_S/\sqrt{\varepsilon_B}$ point is truly the best to maximize the significance of an angular correlation measurement.  We make this choice merely for want of a better guide, and, in particular, lack of reliable information about possible systematic errors.

\subsection{Measurement of azimuthal observables}
\label{subsec:measurement}

In the previous subsection, we investigated six different methods for reconstructing the complete dileptonic \ttbar\ system.  Here, we will now determine to what extent these methods can be used to extract azimuthal correlations.

One issue which we must address at this point is how to define the
production plane in the reconstructed \ttbar\ CM frame, given that the
beams are typically not exactly back-to-back in this frame.  A common
tactic is to follow the Collins-Soper construction~\cite{Collins:1977iv}, which
defines an effective beam axis by taking the difference between the
two unit vectors pointing along the beams.  We take a slightly simpler
route, which we find leads to some improvements in the amplitudes of
our azimuthal modulations.  We simply bring the \ttbar system to rest in
lab coordinates via a rotation-free boost, and then use the lab's
native beam axis for the construction of the production plane. \enlargethispage{-20mm}

\begin{figure}[tp]
\begin{center}
\epsfig{figure=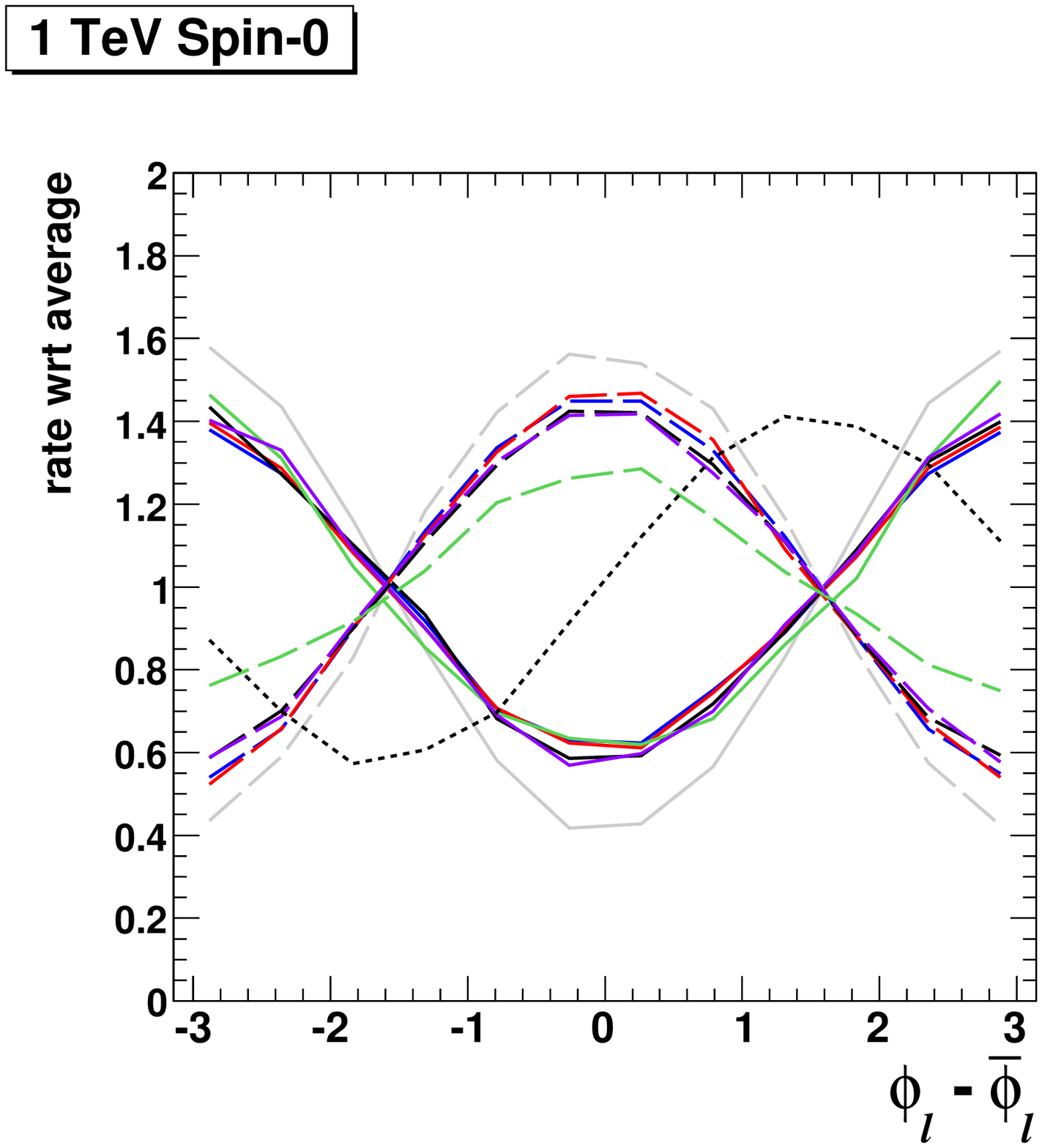,scale=0.40} \hspace{0.0cm}
\epsfig{figure=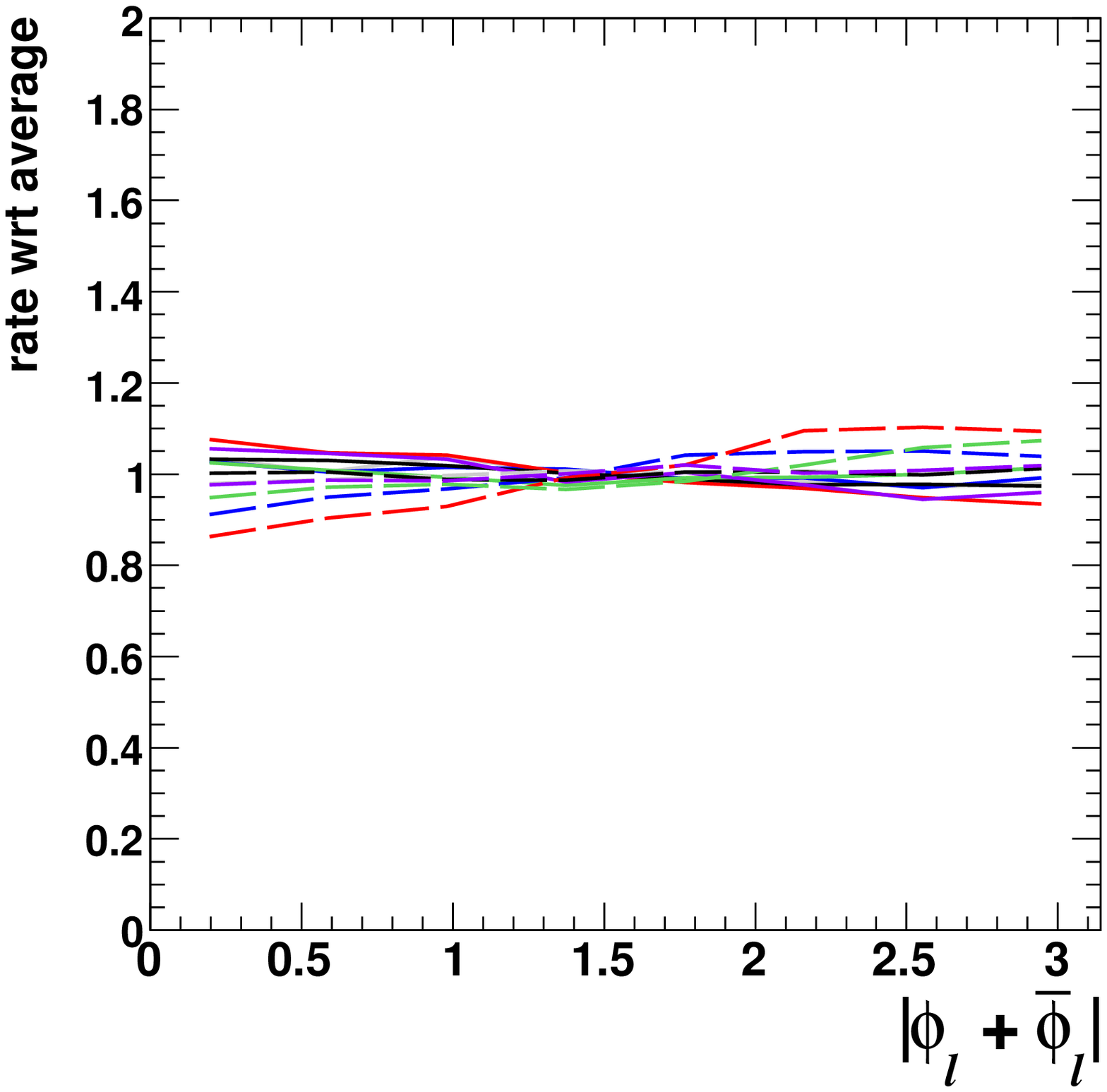,scale=0.40} \hspace{0.5cm}
\epsfig{figure=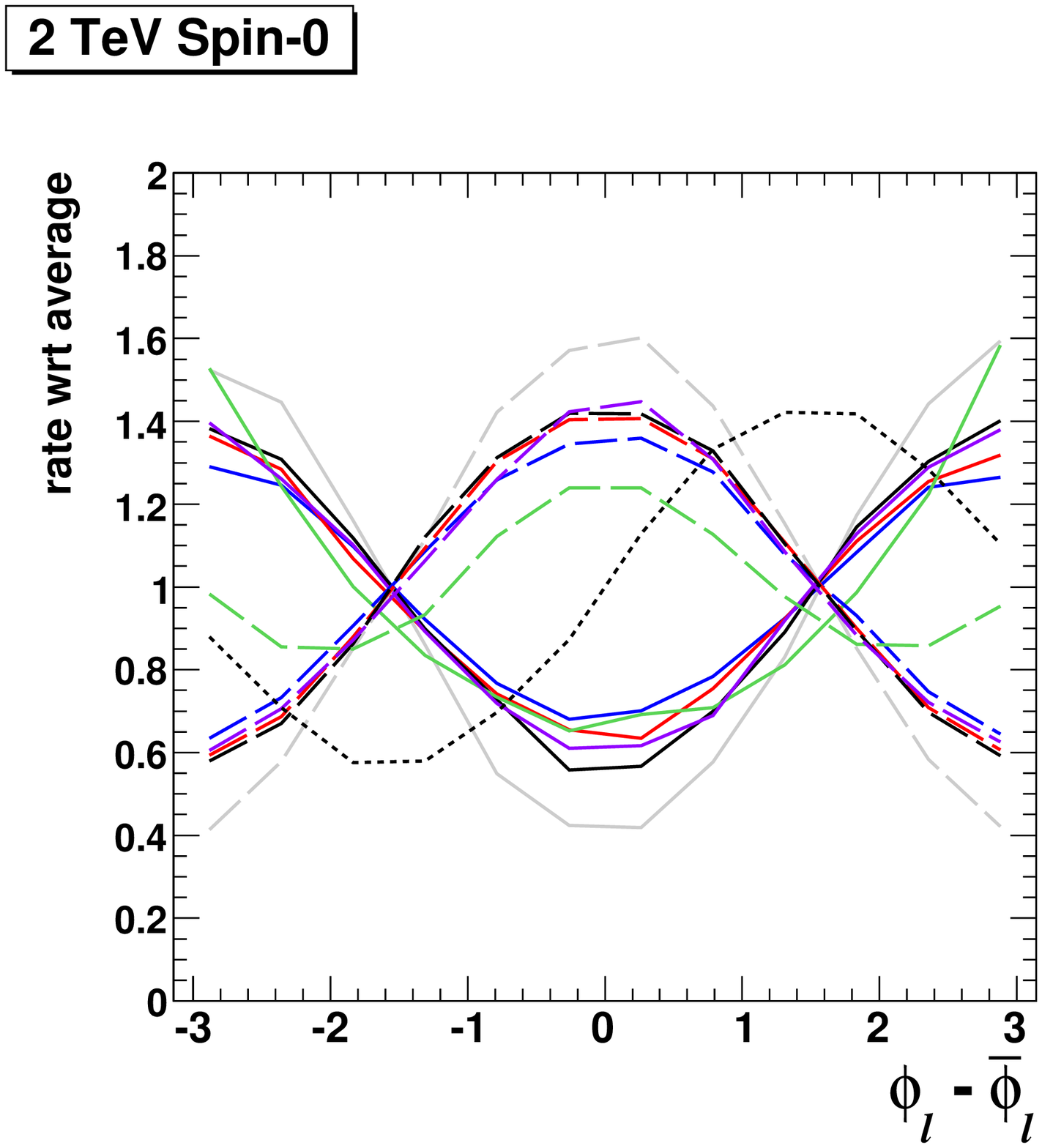,scale=0.40} \hspace{0.0cm}
\epsfig{figure=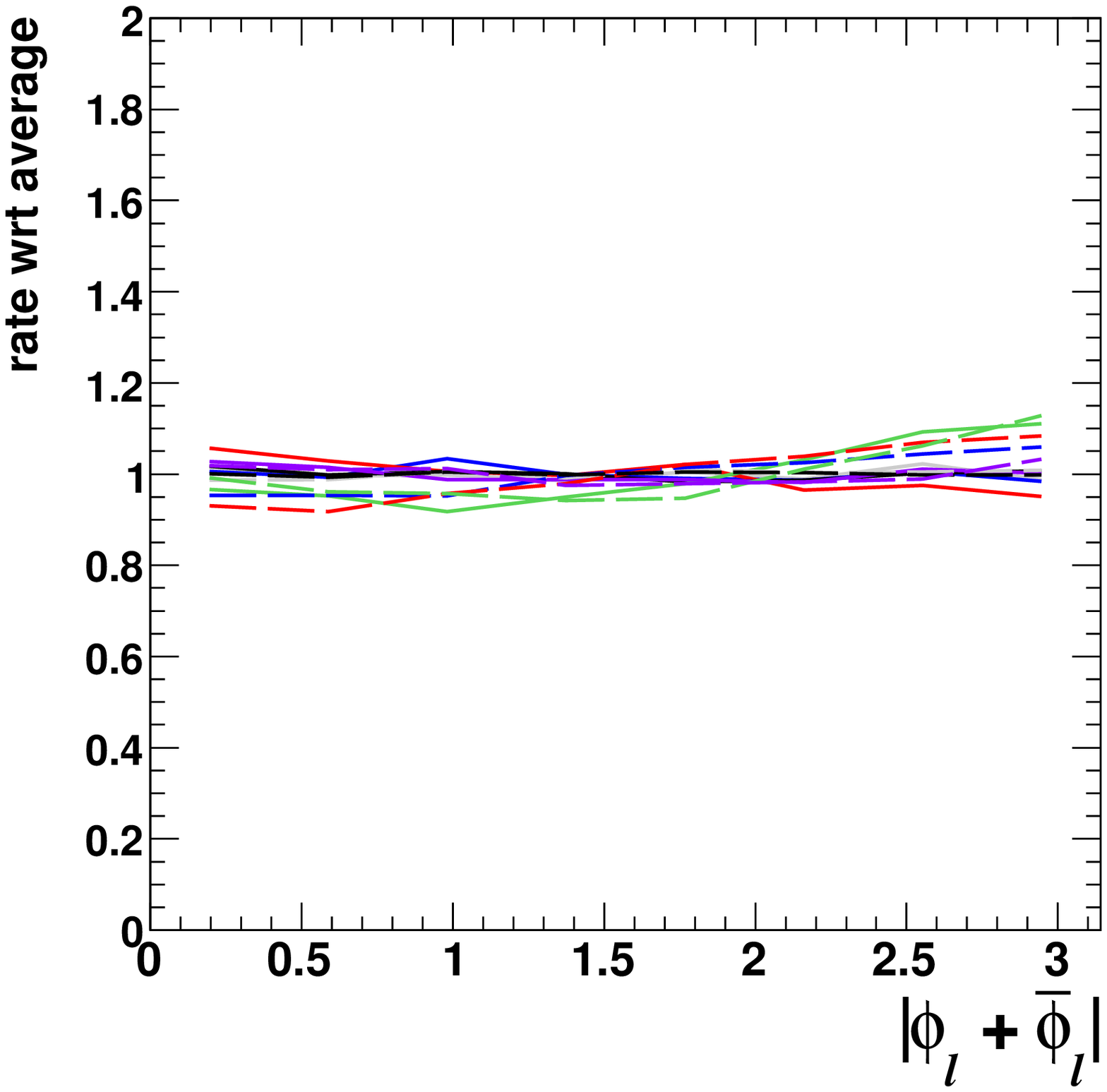,scale=0.40}
\caption{\it Distributions of $(\phi_\ell-\bar\phi_\ell)$ and $|\phi_\ell+\bar\phi_\ell|$ for 1 and 2 TeV spin-0 resonances: pure scalar (solid), pseudoscalar (long-dashed), and mixed $\alpha = \pi/4$ (short-dashed).  Neutrino reconstructions include perfect (grey), $M_{T{\rm cl}}$ (blue), minimal $\eta$-collinear (red), visible-only (black), minimal real-part quartic (green), and Bai-Han quartic (purple).}
\label{fig:spin0modulations}
\end{center}
\end{figure}

\begin{figure}[tp]
\begin{center}
\epsfig{figure=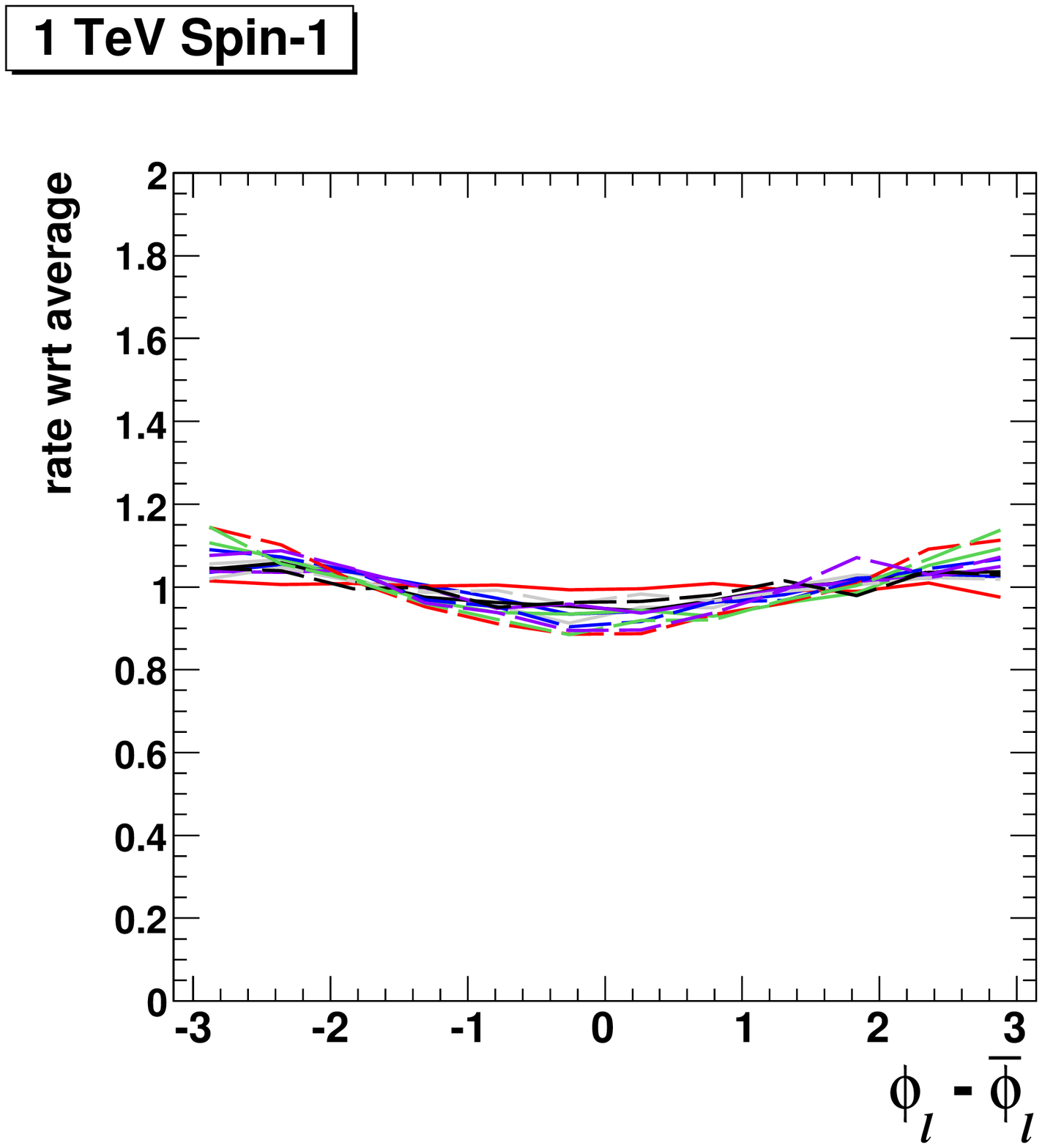,scale=0.40} \hspace{0.0cm}
\epsfig{figure=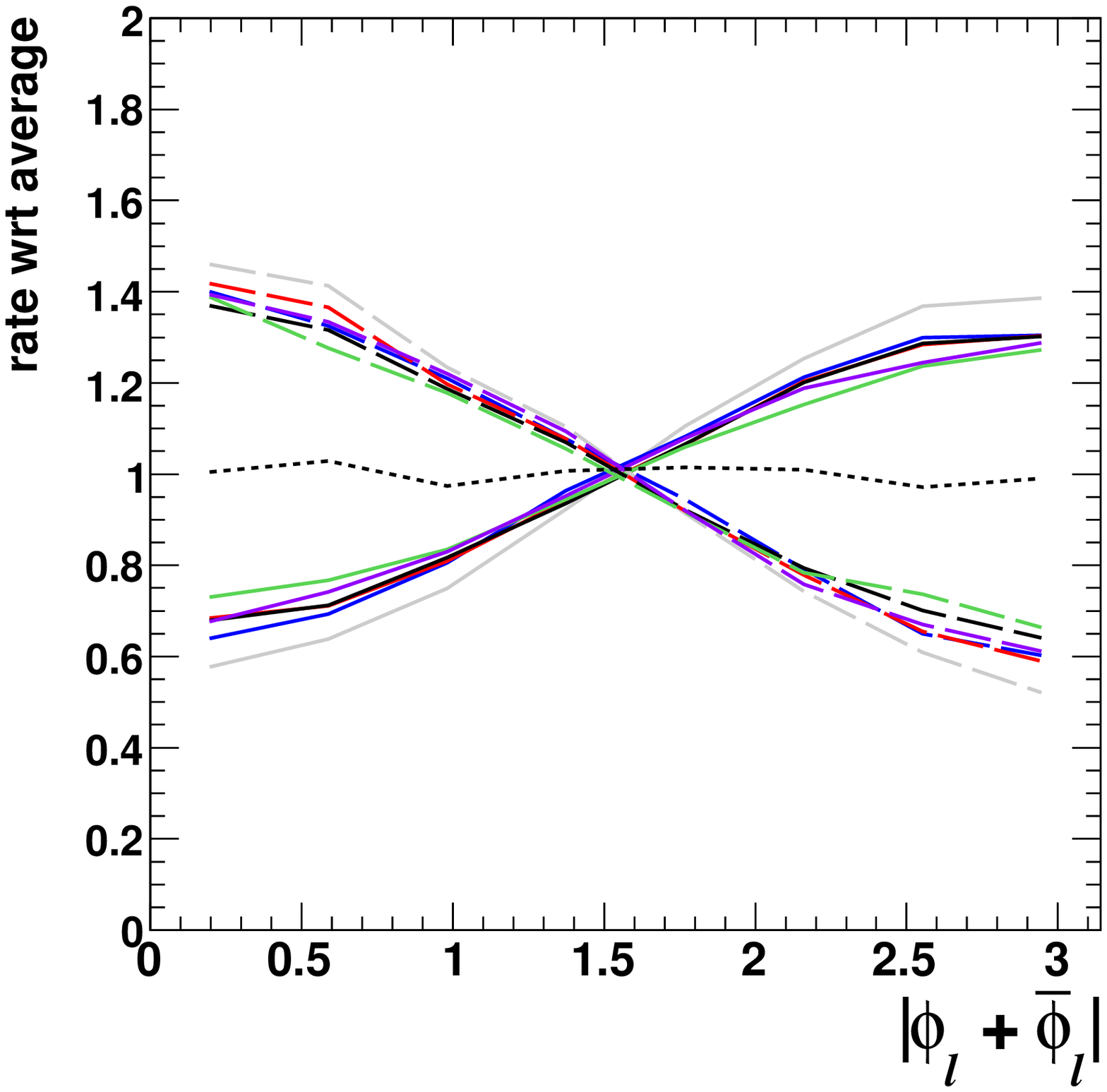,scale=0.40} \hspace{0.5cm}
\epsfig{figure=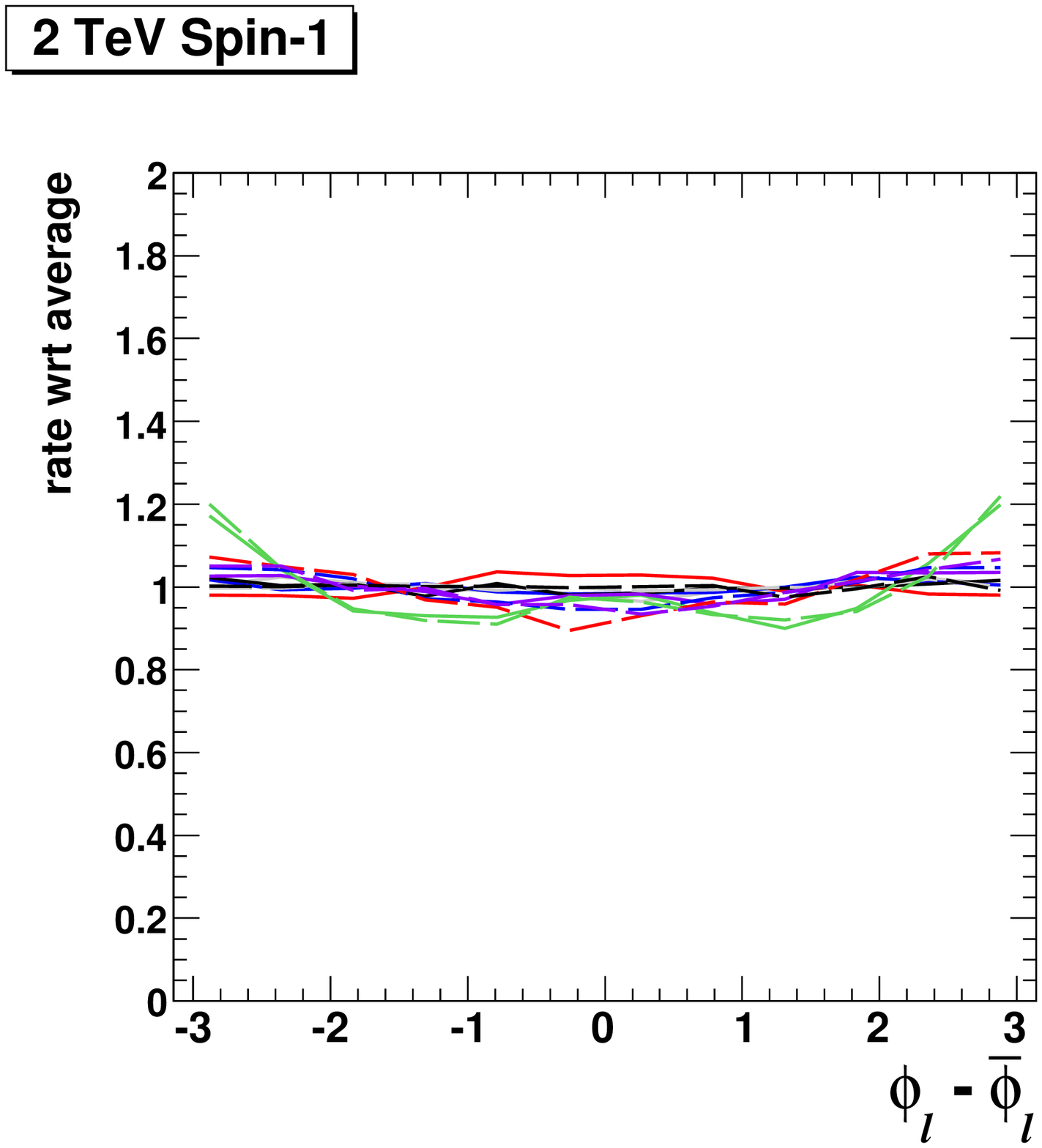,scale=0.40} \hspace{0.0cm}
\epsfig{figure=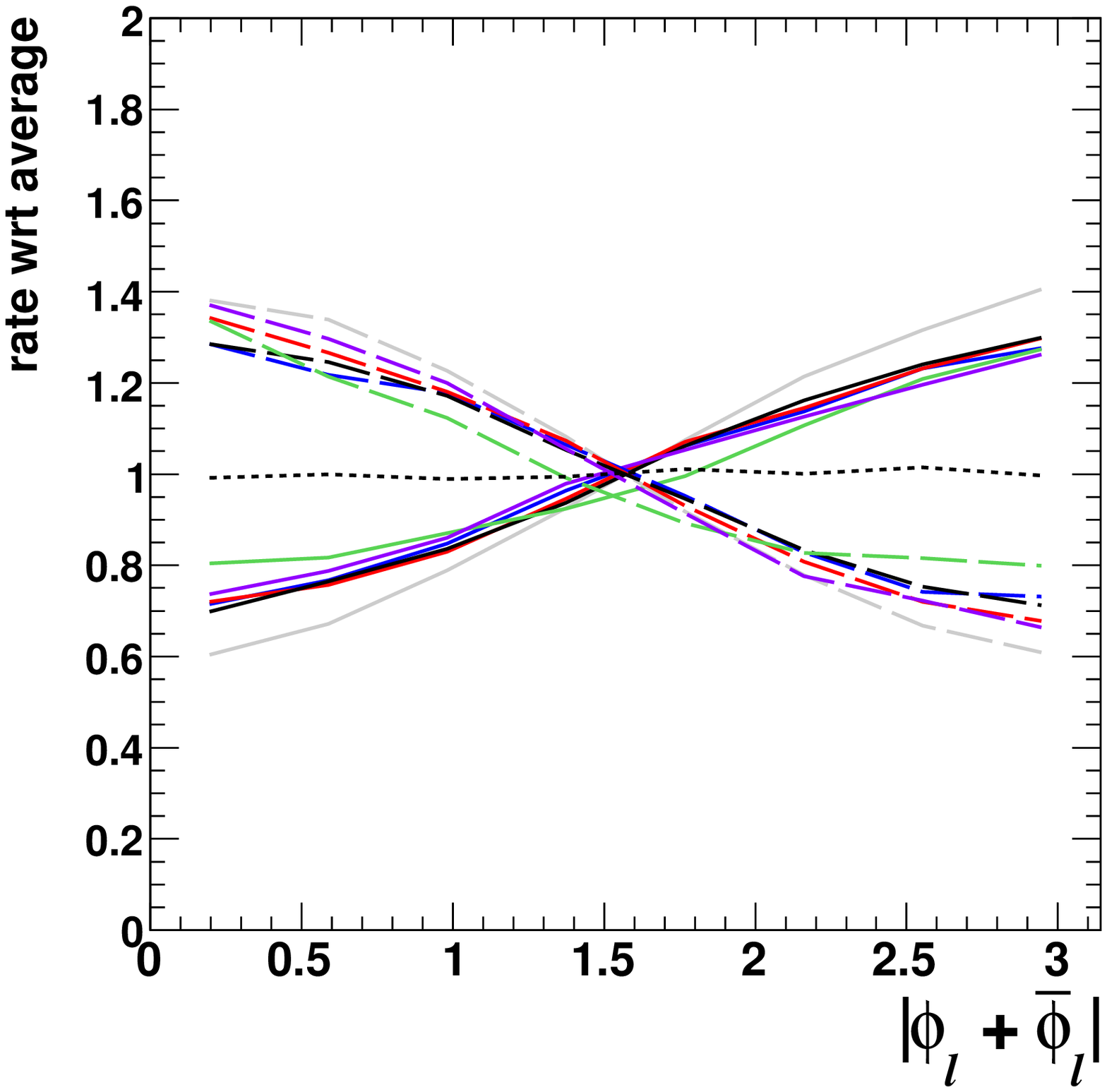,scale=0.40}
\caption{\it Distributions of $(\phi_\ell-\bar\phi_\ell)$ and $|\phi_\ell+\bar\phi_\ell|$ for 1 and 2 TeV spin-1 resonances: pure vector (solid), axial-vector (long-dashed), and left-chirality (short-dashed).  Neutrino reconstructions include perfect (grey), $M_{T{\rm cl}}$ (blue), minimal $\eta$-collinear (red), visible-only (black), minimal real-part quartic (green), and Bai-Han quartic (purple).}
\label{fig:spin1modulations}
\end{center}
\end{figure}

\begin{figure}[tp]
\begin{center}
\epsfig{figure=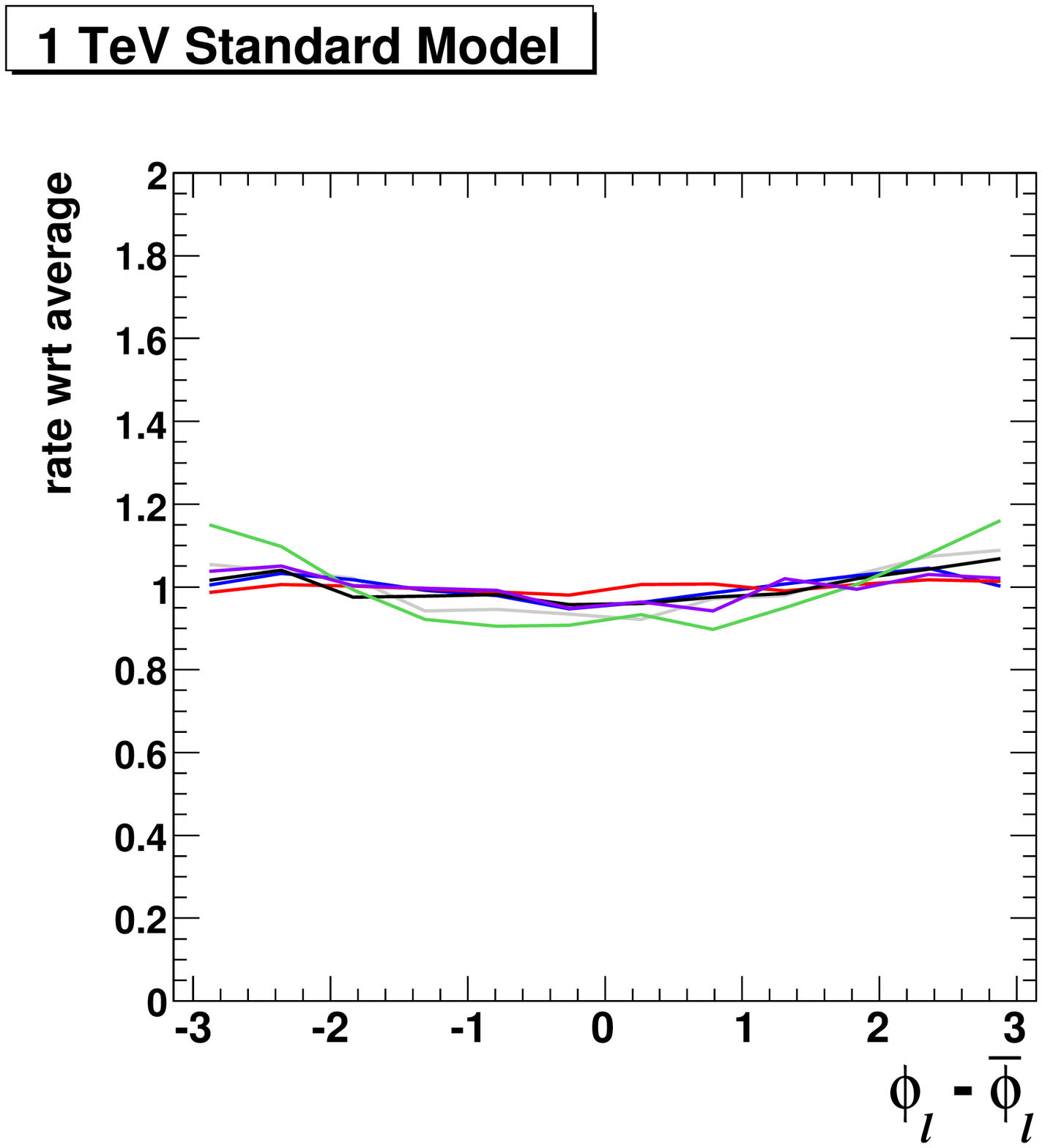,scale=0.40} \hspace{0.0cm}
\epsfig{figure=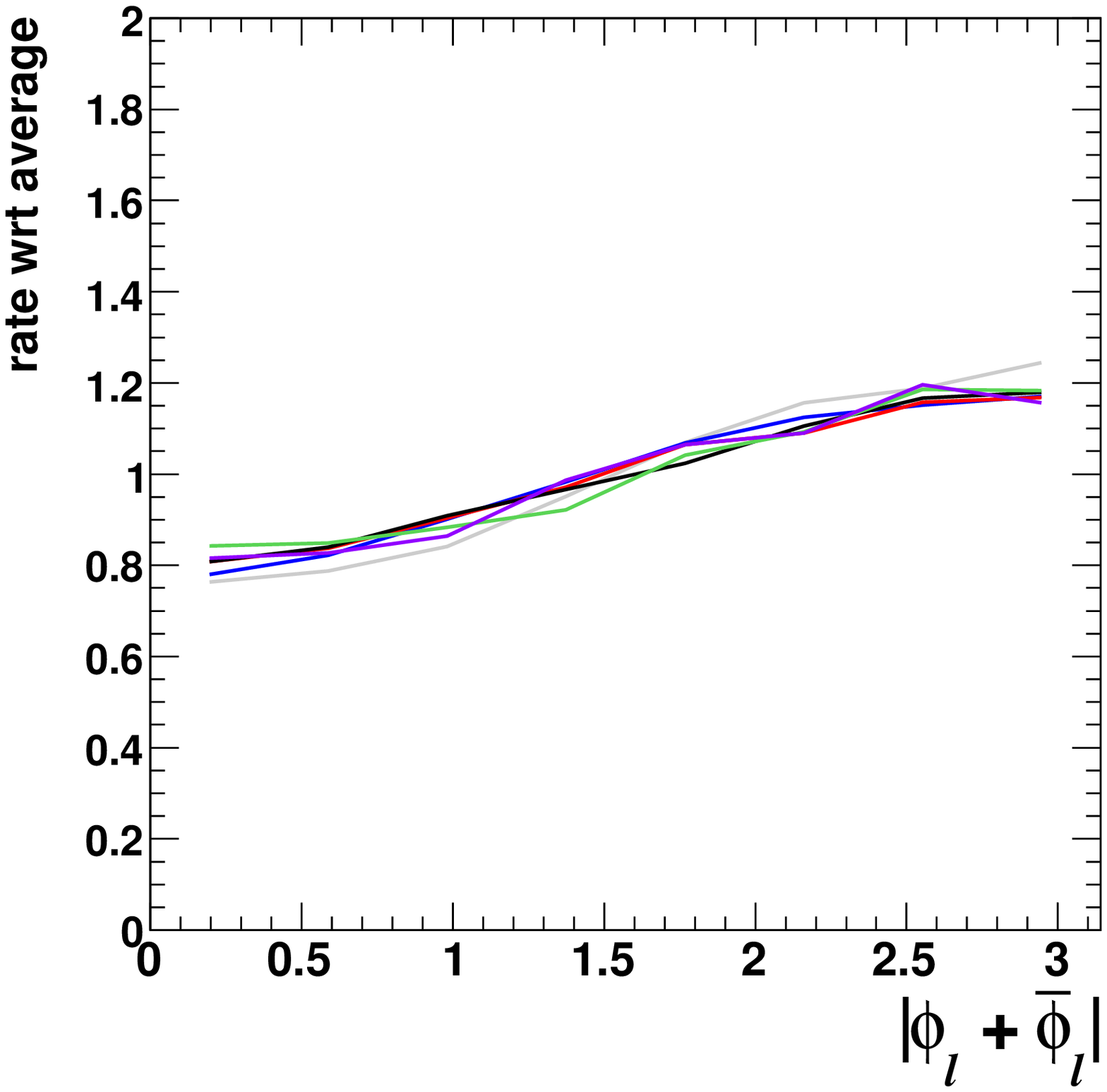,scale=0.40} \hspace{0.5cm}
\epsfig{figure=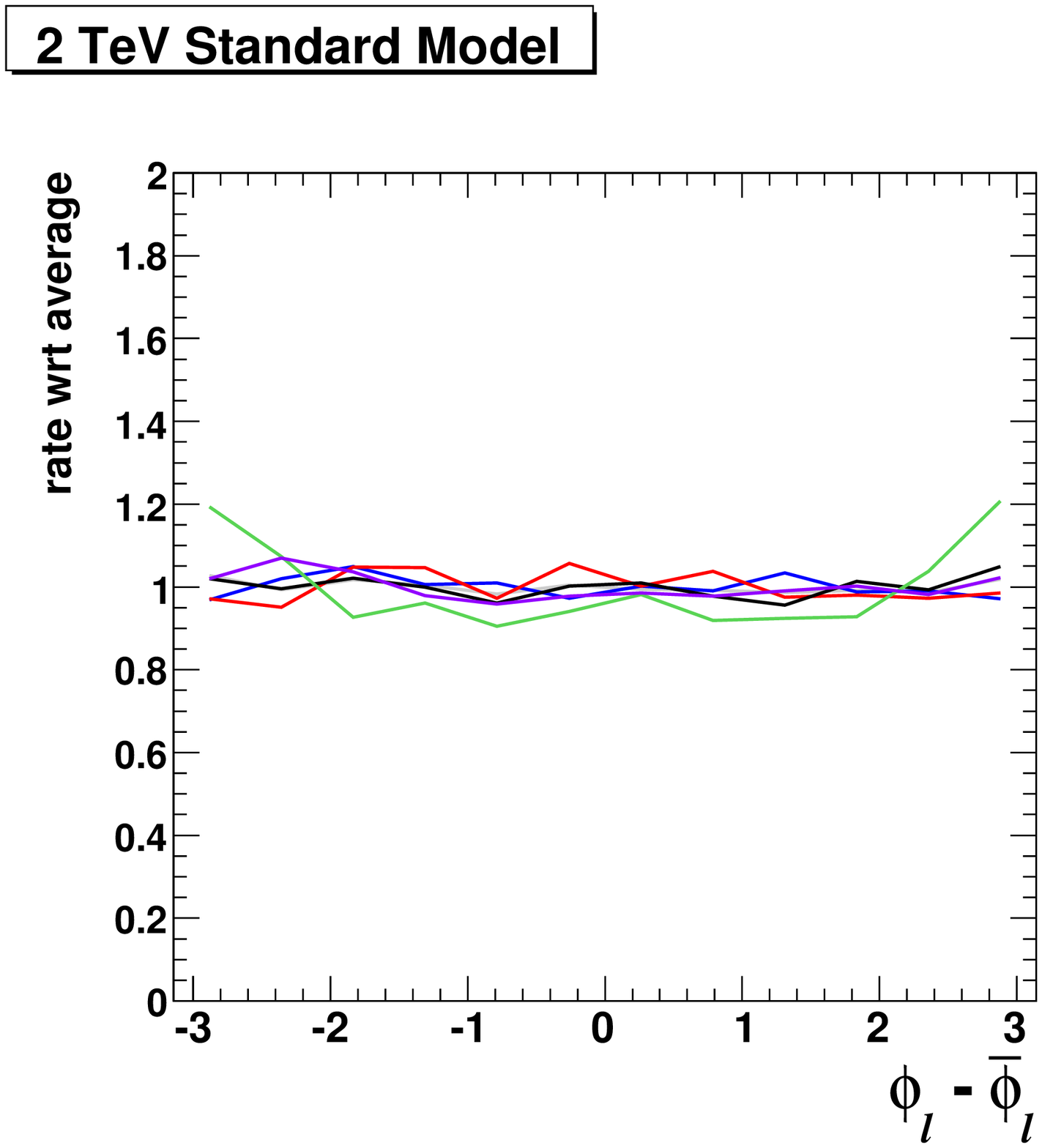,scale=0.40} \hspace{0.0cm}
\epsfig{figure=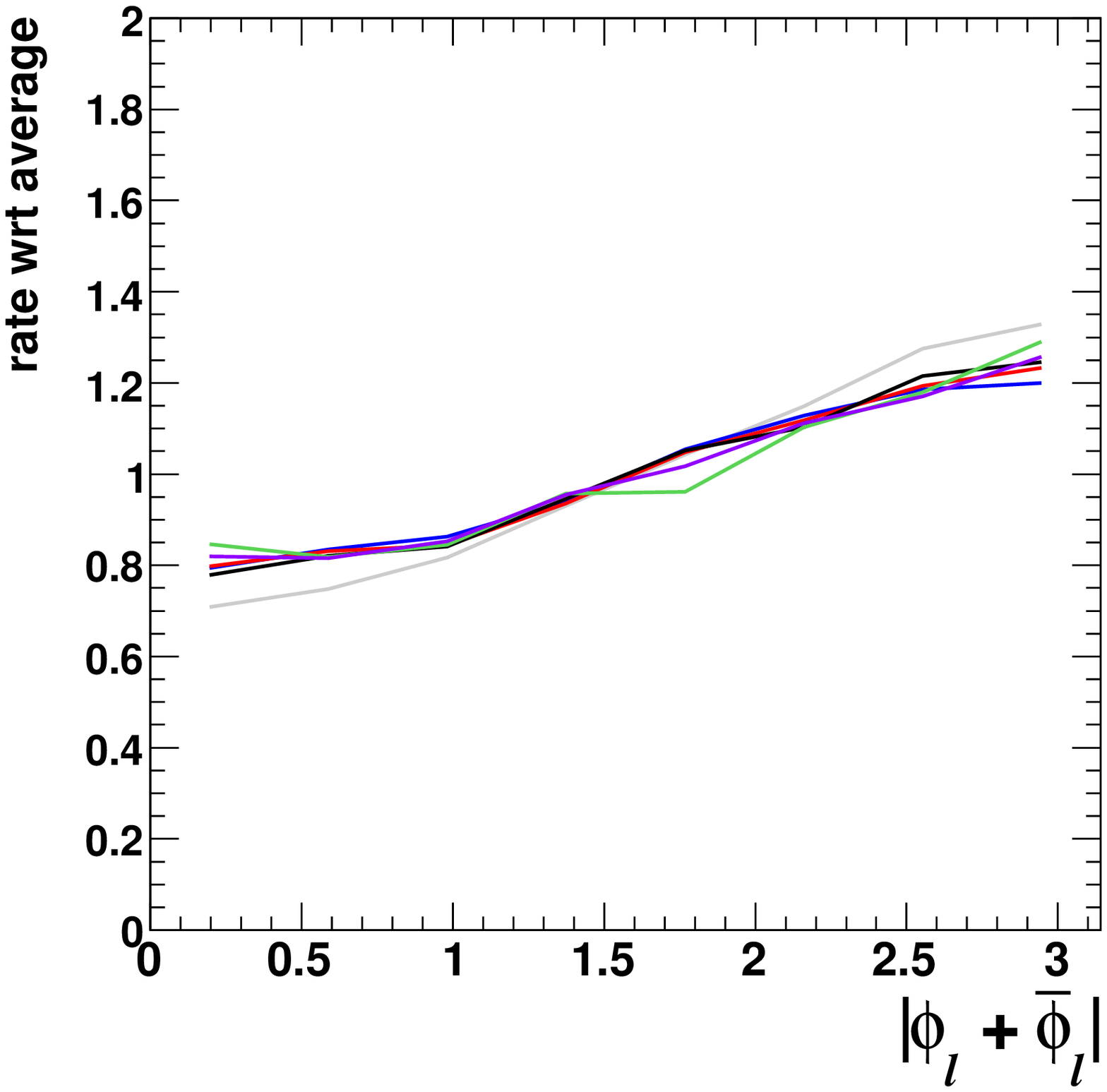,scale=0.40}
\caption{\it Distributions of $(\phi_\ell-\bar\phi_\ell)$ and $|\phi_\ell+\bar\phi_\ell|$ for Standard Model \ttbar\ at LHC14 in 1 and 2 TeV (broad) mass windows.   Neutrino reconstructions include perfect (grey), $M_{T{\rm cl}}$ (blue), minimal $\eta$-collinear neutrino (red), visible-only (black), minimal real-part quartic (green), and Bai-Han quartic (purple).}
\label{fig:SMmodulations}
\end{center}
\end{figure}

In Figs.~\ref{fig:spin0modulations} through~\ref{fig:SMmodulations},
we show the reconstructed distributions of $(\phi_\ell-\bar\phi_\ell)$
and $|\phi_\ell+\bar\phi_\ell|$ for a variety of narrow $gg$-initiated
spin-0 resonances and $q\bar q$-initiated spin-1 resonances, and for
the SM continuum at the 14 TeV LHC.  We apply our narrow resonance
cuts from Table~\ref{tab:eff14} to our resonance samples, and broad
cuts to the SM samples (which slightly improves statistics).  The
reconstructions are the same as those discussed in the previous
subsection, though we now omit $M_{\rm eff}$, which does not offer a
fully 3D picture of the event.  The figures exhibit all of the
expected behavior discussed in Section \ref{sec:spinCorr}, for all of
the reconstruction techniques.\footnote{We do acquire some small
$|\phi_\ell+\bar\phi_\ell|$ modulations for the scalars and
$(\phi_\ell-\bar\phi_\ell)$ modulations for the vectors, up to the
10\% scale for 1~TeV resonances.  These are mainly due to some biases
incurred by our selection cuts and jet-lepton pairing strategy, rather
than genuine helicity interference.  Nonetheless, the effects of these
biases also show some correlations with the resonance couplings.}
Some are clearly better than others, but interestingly the differences
are usually not very dramatic.  The clear exception is the minimal
real-part quartic solution, which tends to introduce spurious
modulations in $(\phi_\ell-\bar\phi_\ell)$.  However, the remaining
reconstructions are all quite good.  In particular, we preserve nearly
the entire modulation using only {\it visible} particles (the two
leptons and two $b$-jets).\footnote{The visible-only $\phi_\ell$'s are
actually very closely related to angular variables which can be
constructed in the top frame, and which are associated with a
well-defined analyzing power of about 80\%: instead of directly
measuring the lepton direction, first boost the $b+l$ composite system
to rest (without rotation), and then measure the lepton direction.  In
our visible construction, we instead immediately boost the $b+l$
system into lab-frame without first bringing it to rest in top-frame,
but the resulting $\phi_\ell$ is nearly the same.  It is even possible
to accurately approximate this using purely geometric constructions in
the lab frame.  For example, each of the $b$-jets can be taken to
define an independent $z$-axis and production plane, with
$\phi_\ell$'s measured with respect to these coordinate systems.  We
thank Stephen Parke for suggesting the possibility of using lab-frame
angles.} \enlargethispage{-20mm} We take this as a strong hint that these angles may be very
straightforward to reconstruct in reality, at least in the case of
somewhat fast-moving tops where the visible products can be cleanly
separated into hemispheres.  The fact that the semileptonic top decay
is roughly 2/3 visible is clearly a tremendous help here, as the
vector-sum of its visible activity gives us an adequate approximation
of the true top direction of flight.  Moreover, our cuts bias us away
from kinematic configurations where the neutrinos are dominating.

A detailed understanding of the observability of these modulations in
the presence of background and systematic uncertainties is beyond the
scope of this paper.  However, we can make some rough estimates for
the timely question of discriminating a true vector from an
axial-vector.  For this purpose, we can think of the $|\phi_\ell +
\bar\phi_\ell|$ distribution with only two bins, i.e.\ as an asymmetry
of roughly $\pm$25\%.  From a purely statistical perspective, with no
background, these two cases could be discriminated at 3$\sigma$ level
if the error on the asymmetry can be brought down to the 15\% level.
This would require less than 100 events.  Given that, after cuts, we
accept only about 20\% of the dileptonic signal, and hence 1\% of the
total \ttbar\ resonance signal, this kind of analysis becomes
statistically possible when the LHC accumulates $O(10,000)$ \ttbar\
pairs from the resonance.  We would therefore require cross sections
at the pb-scale for the LHC7 (assuming 5--10~fb$^{-1}$), and at the
100~fb-scale for the long-term LHC14.  Interestingly, for $\ell$+jets,
which has six times higher branching fraction but 40\% as much
modulation (if we correlate with the $b$-quark on the hadronic side),
the required cross sections would naively not be so different.  On the
other hand, different factors come into play, in particular the need
to identify $b$-(sub)jets and to possibly run a hadronic top-tagger
(as in,
e.g.,~\cite{Brooijmans:2008zza,Thaler:2008ju,Kaplan:2008ie,Almeida:2008yp,Almeida:2010pa,Plehn:2010st,Ellis:2009su,Thaler:2010tr},
and reviewed in~\cite{Abdesselam:2010pt}), but also presumably larger
efficiency and better resonance peak reconstruction.

Of course, if $S/B \lsim 1$, the discrimination becomes more
challenging.  Besides diluting the signal, the Standard Model continuum also exhibits its
own vector-like $(\phi_\ell + \bar\phi_\ell)$ modulation.  Accurate
extraction of the signal modulation would therefore require a 
well-controlled background subtraction.  Assuming this can be done, then models
with $S/B \simeq 1$ could probably be discriminated at 3$\sigma$ with at least a few hundred
dileptonic signal and background events after cuts.  From Table~\ref{tab:eff7}, we see that the SM backgrounds to a 1(2)~TeV analysis after 10~fb$^{-1}$ at the LHC7 would be roughly 200(1) events.  We are background-limited only for resonances at or slightly above 1~TeV, requiring multi-pb cross sections.  Heavier resonances are rate limited.  For 100~fb$^{-1}$ at LHC14, even 2~TeV resonances will be background-limited, requiring low pb-scale for the signal.  While we have not investigated even higher resonance masses, these again quickly become rate-limited at the 14 TeV machine, with 100~fb yielding adequate statistics to discriminate vector from axial-vector in the dileptonic mode.

Given these estimates, what might we expect to learn about
axigluon-like models applicable to the Tevatron top $A_{FB}$ anomaly?
The size of the asymmetry (central value near 50\% for $m_{t\bar t} >
450$~GeV~\cite{Aaltonen:2011kc}) and the lack of major deviations from
the SM prediction for $d\sigma/dm_{t\bar
t}$~\cite{Aaltonen:2009iz,Abazov:2008ny} favor a heavy resonance with
large couplings.  At the same time, LHC searches for dijet resonances
and contact
interactions~\cite{Collaboration:2011aj,Khachatryan:2011as} constrain
the couplings to quarks.  While we do not undertake a full scan of the
parameter space, we can use the analysis of~\cite{Bai:2011ed} to
identify a reasonable example of a phenomenological model, consistent
at 68\%~CL with the Tevatron anomaly and having modest impact on
$d\sigma/dm_{t\bar t}$: a 1.5~TeV resonance with purely axial
couplings to light quarks and top quarks of $g_A^q = (0.7)g_s$ and
$g_A^t = (-3.0)g_s$.\footnote{The contribution to $A_{FB}$ in the
$m_{t\bar t} > 450$~GeV bin would be about 27\% at parton-level.
Combined with the NLO contribution from the Standard Model, this
brings the asymmetry to within 1$\sigma$ of the measured value.}
Assuming that $b_L$ also couples with the same strength as the tops,
the width is about 18\%, and the branching fraction to tops is close
to 57\% (to light quarks it is about 14\%).  This resonance easily
evades the exclusion limits on universal axigluons and on more general
dijet resonances in the recent ATLAS
search~\cite{Collaboration:2011aj}, is safe from quark contact
interaction constraints,\footnote{We have explicitly checked that the
contribution to central dijet production in each of ATLAS's mass bins,
including full interference and $t$-channel exchanges, is below the
stated uncertainties.} and is below the current CMS limit on top
resonances~\cite{CMS:2011tt}.  At LHC7, $\sigma\times BR(t\bar t)$ is
about 4~pb, and the optimal \mtcl\ window should be highly signal
dominated.  A 5--10~fb$^{-1}$ data set should therefore be enough to
establish a pure axial interpretation over a pure vector
interpretation to at least 3$\sigma$ for this model.  Clearly, then,
azimuthal decay correlations could potentially have a nontrivial part
to play in disentangling the full story behind the Tevatron $A_{FB}$
anomaly, even for the early phase LHC.

It is also worth noting that our analysis could still apply to cases without a well-defined resonance peak.  For axigluon-like resonances well above 1 TeV, matching the size of the Tevatron anomaly requires large couplings, and therefore typically quite large width ($\Gamma/M \gsim 20\%$).  This may simply show up as a broad excess over the Standard Model at high mass.  However, neglecting for the moment interference with the Standard Model, none of our results were actually sensitive to {\it where} we looked on the resonance's Breit-Wigner curve.  Indeed, while a broad resonance signal may be difficult to isolate on-peak, it can come to dominate the high-mass tail of \ttbar\ production.  With a full 14 TeV LHC at $O$(100)~fb$^{-1}$ luminosity, this tail will be very well-populated by the Standard Model alone.  For example, the region above 2 TeV should contain $O$(10,000) central QCD-induced events, and this number may be substantially enhanced by the resonance.

Still, the final fate of such an analysis becomes more
model-dependent, as interference with the Standard Model could be
non-negligible.  This affects not only the total rate, but also the
coefficient of $\cos(\phi_\ell + \bar\phi_\ell)$, in proportion to the
product of the light quark and top quark vector charges.  Notably,
this interference is washed-out if either the top quarks or the light quarks
are mostly axially coupled.  See Appendix~\ref{app:spin1} for a more detailed discussion.

\section{Conclusions}  \label{sec:conclusions}

The discovery of a resonance in the \ttbar\ invariant mass spectrum at the LHC would represent the beginning of a major advance in our understanding of TeV-scale physics.  In the event of such a discovery, determining the precise nature of this resonance will become one of the highest priorities of the LHC experimental program.  Fortunately, the very distinctive decays of top quarks will give us the opportunity to address rather detailed questions about the resonance's couplings.  Unlike the case of a resonance decaying directly into stable particles such as electrons, different \ttbar\ helicity channels coherently interfere as they decay into the same set of 6-body final-states.  The final decay distributions, correlated between top and antitop, therefore carry information about signs/phases in the tops' chiral production matrix elements.

We have argued in this paper that the tops' azimuthal decay
distributions about their production axis provide a sensitive and
experimentally robust probe of these correlations.  As is usual for
\ttbar\ spin correlations, the charged lepton (or down-type quark)
exhibits the largest effects.  Spin-0 resonances lead to 60\%
modulations in $(\phi_\ell-\bar\phi_\ell)$, with a phase offset that
directly represents the resonance's CP phase.  Spin-1 and spin-2
resonances lead to modulations in $(\phi_\ell+\bar\phi_\ell)$, with a
signed magnitude that depends on the ratio of chiral couplings,
$(30\%)\times 2\,/\left(g_L/g_R+g_R/g_L\right)$ when integrated over
production angles (two times larger when restricting to highly central
production).  While the spin-0 CP phase has long been a topic of
investigation, the spin-1(2) chiral coupling ratio is largely missed
in common top spin correlation variables such as the double-cosine
distribution and the 3D lepton opening angle.  We have found that
single-differential azimuthal sum/difference distributions contain all
of the relevant information contained in these more common variables,
and more.

We have also found that these azimuthal distributions are straightforward to reconstruct, without detailed information on neutrino kinematics.  We established this point in the dileptonic mode, in which variables sensitive to spin correlations usually require complicated reconstruction techniques.  In contrast, azimuthal distributions can be well-approximated using only visible kinematics.  The major nontrivial task then becomes isolating the resonance region by reconstructing \mtt.  Our own (admittedly limited) investigation suggests that simple reconstructions such as \mtcl\ may be superior to more complete reconstructions that utilize on-shellness criteria on the $W$s and tops.  However, this conclusion could still change in a more detailed experimental study, or with better reconstruction techniques.

Since the modulations are large in the dileptonic mode, measuring them
should not require a very big event sample.  For example, $O(100)$
events should be adequate to discriminate a pure vector resonance from
an axial-vector resonance at better than 3$\sigma$ level, assuming
$S/B \gsim 1$.  For favorable choices of parameters, a color-octet
axial-vector resonance could explain the Tevatron $A_{FB}$ anomaly, be
discovered at the early LHC, {\it and} have its mostly-axial coupling
to tops confirmed not long afterwards.  This would be a dramatic
interplay between the capabilities of the two colliders.

While we have focused on dileptonic \ttbar,\ the option of using the \ljets\ mode will also be important to explore in more detail.  Extracting the smaller modulations may require greater control over systematics, but the much higher statistics and much better kinematic reconstruction afforded by this channel could easily offset this.  In particular, for resonances well above 2 TeV, \ljets\ may be the only viable option, even for the long-term LHC, due to the smaller cross sections involved.

We have also limited ourselves to simulations based on leading-order
\ttbar\ production and decay, with additional radiation and kinematic
rearrangements provided by the \PYTHIA\ virtuality-ordered parton
shower.  Understanding the genuine NLO behavior of the azimuthal
correlations, processed through realistic event reconstructions,
remains as an interesting open task which can potentially be addressed
by modern
calculations~\cite{Bernreuther:2010ny,Melnikov:2009dn,Melnikov:2010iu}.

Finally, while the focus of this work has been characterization of physics beyond the Standard Model, we note that azimuthal decay angles could also be useful for detecting the spin correlations of boosted or semi-boosted tops from pure QCD ({\it cf.} \hspace{-3mm} the modulations in Fig.~\ref{fig:SMmodulations}), and serve as a particularly clean indication that the top quark is spin-1/2.


\acknowledgments{We thank Xiaozhou Zhou for collaboration in the early
stage of this work.  We also thank Zhenyu Han and Kirill Melnikov for
discussions, and Andrei Gritsan for comments on the manuscript.  MB
thanks Boston University for accommodation during portions of this
work.  BT thanks the Berkeley Center for Theoretical Physics and the
University of Oregon Institute of Theoretical Science for their
hospitality.  MB was supported by NSF grant nsf-phy/0910467.  BT was
supported by DoE grant No.\ DE-FG-02-91ER40676.}


\appendix

\section{Basic Definitions and Spin-0}
\label{app:spin0}

In Section~\ref{sec:spinCorr}, we gave various formulas for $d\Gamma$ over 
different portions of the angular phase
space, including the production angle of the $t \bar t$ pair in the  \ttbar\ CM
frame and the decay angles for the two leptons.  In these appendices, we present
these formulas in a more complete form, displaying all angular correlations.
We also go beyond the chiral $m_t=0$ limit.  Though
this is likely to be a reasonable approximation for resonances at 1~TeV or higher,
finite-mass effects are not always totally negligible.  We parametrize them with
factors of $r \equiv 2m_t/M$ (the inverse Lorentz boost of the tops) and
$\sqrt{1-r^2}$ (the tops' velocities).  However, we continue to ignore 
velocity-suppression factors that arise purely from \ttbar\ phase space, which in any
case can be factorized out of the total resonance decay rate. 
We also neglect interference terms with the SM.  These effects tend to be small on-peak.
There are, of course, cases where the interference is nonetheless more important than
the finite-mass corrections.  We outline out the effects of interference in more detail below.

Recalling Section~\ref{sec:spinCorr}, we define our coordinate system as follows.  
We first find the $t \bar t$ production axis in the \ttbar\ CM frame.  The angle that this
makes with the beam axis is called $\Theta$, and we average over configurations
$\Theta \leftrightarrow \pi-\Theta$.  We then perform
rotation-free boosts to bring both tops to rest.  We define a common
$z$-axis using the original $t$ direction.  To define the $y$-axis, we
take either of the two unit vectors perpendicular to $t \bar t$ production plane, 
and then define the $x$-axis to make a right-handed 3D coordinate system.

Before proceeding, we note that all of the
equations that we give can apply to arbitrary pairs of top decay products.  
For example, in the \ljets\ channel we might correlate the lepton with the
$b$-jet from the hadronic side.  
Starting with the full angular distributions for two leptons, upon changing to the 
angular variables for the particles of interest, we need only multiply 
each $\sin$ and $\cos$ of the new polar angle by the appropriate spin-analyzing 
power: 1 for leptons, -0.3 for neutrinos, -0.4 (+0.4) for 
$b$-quarks ($W$-bosons), and 0.5 for the softer of the two
$W$ decay products in the top rest frame.\footnote{Note that care should be taken if, for example, we decide instead to define the $z$-axis by following the direction of the semileptonic top in \ljets.  In that case, the formulas for the $\ell^+$+jets events stay the same, but for the $\ell^-$+jets events they require replacements $\theta\to\pi-\theta$ and $\bar\theta\to\pi-\bar\theta$.}$^,$\footnote{The analyzing powers for particles other than the lepton are only formally correct under the assumption that the phase space of the remaining particles is fully integrated.  In principle, they could change after accounting for detector acceptance and analysis cuts.}
We subsequently drop the ``$\ell$" subscript from all of the angular variables, as an
implicit reminder that we are no longer restricted to dileptonic mode.

The angular formulas for spin-0 decay were given in
Section \ref{subsec:spin0} to zeroth order in $r$.  
Restoring the full kinematic dependence, we get
\be
\frac{d^4\Gamma}{d\Omega \, d\bar\Omega} &\propto& 
1 + \cos\theta \cos\bar\theta 
- \sin\theta \sin\bar\theta \Big(\! \cos(2 \alpha) \cos \left(\phi - \bar\phi \right) \nonumber \\
&& \hspace{5cm} - \sqrt{1-r^2} \sin (2 \alpha ) \sin (\phi-\bar\phi) \Big) \nonumber \\
&&- \frac{1}{2} r^2 \big(1+\cos (2 \alpha ) \big)
   \big( 1+\cos\theta \cos\bar\theta- \sin\theta \sin\bar\theta \cos (\phi-\bar\phi) \big), \nonumber \\
\frac{d\Gamma}{d (\phi-\bar\phi)} &\propto& \Big( 1 - r^2 \cos^2\alpha  \Big) \nonumber \\
&-& \left( \frac\pi4 \right)^2 \Big( \big[ \cos(2\alpha) - r^2 \cos^2\alpha \big]\cos(\phi - \bar\phi) - \sqrt{1-r^2} \sin(2\alpha) \sin(\phi-\bar\phi)  \Big),
\label{eq:appSpin0}
\ee
where $2\alpha$ is the relative phase between the $\bar{t}_L t_R$ and
$\bar{t}_R t_L$ couplings.  Values of $\alpha$ besides 0 and $\pi/2$
signify CP violation.  The upper formula is the ``master'' leptonic
angular distribution from which one can derive the dependence on
whatever angular variable one wishes.  The lower equation describes the
azimuthal correlations, which we found useful for identifying a heavy
particle's relative couplings to different top chiralities regardless
of its spin.  In this case, the dependence is in $(\phi - \bar\phi)$.
For spin-0, $(\phi+\bar\phi)$ is a physically meaningless quantity,
though we will see below that spins 1 and 2 have their leading
modulations in it.  Formulas for the traditional distributions in
$\cos \theta \cos\bar\theta$ and $\cos\chi$ are given in Section
\ref{sec:spinCorr}.

In addition to the finite-boost corrections, we can also consider the effects of interference with
the Standard Model.  First, we note that these effects vanish for a scalar produced in $q\bar q$ annihilation,
since to very good approximation the QCD contribution only takes place between quarks of opposite helicity
(same spin).  For $gg$ fusion, the interference vanishes at high boost, as the QCD contribution dominantly 
proceeds from same-spin gluons to same-spin (chiral) tops.  The interference is also suppressed relative to
the ``resonance-squared'' contribution since the interference term crosses a zero on-peak.  
Integrating the interference contribution across the peak (accounting for imperfect cancellation due to falling PDFs and/or
asymmetric signal window), the total relative suppression goes as $r(\Gamma/M)^2$.  In the case
of a color-singlet scalar with generic couplings to $G^{a\mu\nu} G^a_{\mu\nu}$ and $G^{a\mu\nu}\tilde{G}^a_{\mu\nu}$,
the interference in the $(\phi-\bar\phi)$ modulation is controlled by the latter coupling for production nearly
perpendicular to the beamline (and swamped by the quasi-singular pure QCD rate for production near the beamline).  
While the resonance-squared contribution exhibits a modulation $\cos(\phi-\bar\phi+2\alpha)$, the interference comes 
in with $\sin(\phi-\bar\phi+\alpha)$, for example signalling $CP$-violation when the resonance couples to tops as a pure
scalar ($\alpha=0$) and to gluons as a pure pseudoscalar.  For a 1~TeV resonance of 20\% width, we estimate that the size of the interference
contribution relative to the resonance-squared is less than 20\% if $S/B > 1/4$.  This relative contribution decreases with increasing mass, decreasing $\Gamma/M$, or increasing $S/B$.\footnote{Quite
generally, we can infer that the relative contribution of interference within a mass window of width $O(\Gamma)$ centered on the resonance is at most $(\Gamma/M)/\sqrt{S/B}$, with a
typically $O$(1) coefficient that depends on the specific process.  In our case, this coefficient carries an additional suppression factor
of $r$.}

\section{Spin-1}
\label{app:spin1}
   
We give the tops' coupling to a spin-1 particle in Eq.~\ref{eq:spin1Lag}.  
Since the interactions involve same-chirality fields, the left- and right-handed
couplings are independent in the absence of parity. We parametrize the different
chiral couplings as follows,
\be
g_L &=& g \cos\xi \nonumber \\ 
g_R &=& g \sin\xi.  \label{eq:xiDef}
\ee
Just as with the scalar case, including $m_t \neq 0$ in the resonance decay will modify
the differential decay rate.  Since the equivalence
of chirality and helicity breaks down with finite fermion mass, all possible
spin pairings contribute, though only $\uparrow \uparrow$ 
and $\downarrow \downarrow$ remain in the limit $r \rightarrow 0$.

As mentioned above, producing a spin-1 resonance with zero angular momentum along
the beam axis requires gluon fusion, which only occurs via a higher-dimension
operator.  We therefore begin with formulas for the likely more dominant scenario
of $J_{\rm beam}=\pm1$ via $q \bar q$ annihilation.  

Including the dependence on the production angle and all decay angles (choosing one 
particle each from the $t$ side and $\bar t$ side), we get
\be
\frac{d^5\Gamma_{J_{\rm beam}=\pm1}}{d\Omega \, d\bar\Omega \, d\cos\Theta} &\propto& 
 \big( 1 + \cos^2\Theta \big) \Big( 1 - \cos\theta\cos\bar\theta  + \sqrt{1-r^2} \cos(2\xi) 
 \big(\cos\bar\theta - \cos\theta \big) \Big) \nonumber \\
&&- \sin(2\xi)\sin^2\Theta\sin\theta\sin\bar\theta\cos(\phi+\bar\phi) \nonumber \\
&+& \frac12 r \Big[ \big( 1 + \sin(2\xi) \big) \sin(2\Theta)  \big( \cos\theta\sin\bar\theta\cos\bar\phi + \cos\bar\theta\sin\theta\cos\phi \big) \nonumber \\
&& \hspace{1cm} + \sqrt{1-r^2} \cos(2\xi) \sin(2\Theta)  \left(\sin\theta\cos\phi - \sin\bar\theta\cos\bar\phi \right) \Big] \nonumber \\
&-& r^2 \Big[\cos^2\Theta - \sin(2\xi) + \big(\sin(2\xi)\cos^2\Theta -1\big)\cos\theta\cos\bar\theta  \nonumber \\
&& \hspace{1cm} +  \sin^2\Theta \sin\theta \sin\bar\theta \lp \cos\phi \cos\bar\phi + \sin (2\xi) \sin\phi \sin\bar\phi \big) \Big].
\label{eq:niSpin1Pol1}
\ee
From this distribution, we can
integrate and/or change variables to get a function of whatever
angular variables we want.  We start by demonstrating the claim made in Section
\ref{sec:spinCorr} that for a sufficiently light resonance, one can
determine the extent to which it is vector or axial with the
traditional $\cos\theta\cos\bar\theta$
distribution.  We can see the difference by setting $\xi = \pi/4$ or
$3\pi/4$ and integrating the other angles in Eq.~\ref{eq:niSpin1Pol1}.  
Dividing out by a common numerical prefactor, we get
\be
\frac{d\Gamma_V}{d\cos\theta \, d\cos\bar\theta} &\propto& \left( 1 + \frac{r^2}{2}\right) 
- \left(1 - \frac{r^2}{2} \right) \cos\theta\cos\bar\theta \nonumber \\
\frac{d\Gamma_A}{d\cos\theta \, d\cos\bar\theta} &\propto& (1-r^2) \big(1 - \cos\theta\cos\bar\theta \big).
\label{eq:finiteMtSpin1}
\ee
Firstly, we note that this reproduces the values listed in Table II of
Ref.~\cite{Frederix:2007gi}.  We see that the ratio of modulating to
constant term for the vector only approaches that of an axial
resonance in the chiral limit.  This difference at nonzero $r$
allows one to separate the two cases, though for $M$ above 1~TeV, 
this requires fitting the 2D distribution to percent-scale precision.  
Lastly, we note that the decay amplitude vanishes for an axial-vector at \ttbar\ threshold.
This is simple to understand with parity. 
An axial vector is even under $P$, while the exchange of $t \bar t$ in the final state
is odd.  For the fermion pair at rest, there is no orbital contribution to the
wavefunction that can make up the difference, so the amplitude shuts off.
We give the distributions in $(\phi + \bar\phi),\, (\phi - \bar\phi)$ in Eq.\ref{eq:appAzCorrOnly0}.

For completeness, we include the analogue of Eq.~\ref{eq:niSpin1Pol1} 
for $J_{\rm beam}=0$, as would arise for a color-octet produced in gluon fusion.
Production of such a state is usually suppressed due to the need for higher-dimension interactions, but could dominate
if dimension-four couplings to light quarks are small.  
The 5D distribution in leptonic and production angles is
\be
\frac{d^5\Gamma_{J_{\rm beam}=0}}{d\Omega \, d\bar\Omega \, d\cos\Theta} &\propto&
\sin^2\Theta \Big(1-\cos\theta\cos\bar\theta + \sin(2\xi) \sin\theta\sin\bar\theta \cos(\phi+\bar\phi) \nonumber \\
&& \hspace{1.5cm} + \frac14 \sqrt{1-r^2} \cos(2\xi) \big(\! \cos\bar\theta - \cos\theta \big) \Big) \nonumber \\
&-& \frac18\, r \Big[ \big(\! 1+\sin(2\xi)\big) \sin (2 \Theta)  \big(\! \cos\theta\sin\bar\theta\cos\bar\phi + \cos\bar\theta\sin\theta\cos\phi \big) \nonumber \\
&& \hspace{1cm} - \sqrt{1-r^2} \cos(2\xi) \big(\! \sin\bar\theta\cos\bar\phi - \sin\theta\cos\phi \big) \Big] \nonumber \\
&+& \frac12\, r^2 \Big[\big(\! 1+\cos(2\Theta)\sin(2\xi) \big)  \big( \cos\theta\cos\bar\theta - \sin\theta\sin\bar\theta\sin\phi \sin\bar\phi \big) \nonumber \\
&& \hspace{1cm} + \big(\!\cos(2\Theta)+\sin(2\xi)\big) \big(\! 1 - \sin\theta\sin\bar\theta\cos\phi\cos\bar\phi \big) \Big].
\label{eq:niSpin1Pol0}
\ee
From Eqs.~\ref{eq:niSpin1Pol1} and \ref{eq:niSpin1Pol0}, we derive azimuthal distributions for polarizations $\pm1$ and 0,
\be
\frac{d\Gamma_{J_{\rm beam}=P}}{d (\phi + \bar\phi)} &\propto& 
   \left(1 + \frac14 r^2 \big(3\sin(2\xi)-1\big)\right) - \zeta_P \left(\frac{\pi}{4}\right)^2 \left( \sin(2\xi) + \frac12 r^2 \big( 1-\sin(2\xi) \big) \right)  \cos(\phi+\bar\phi)  \nonumber \\
\frac{d\Gamma_{J_{\rm beam}=P}}{d(\phi - \bar\phi)} &\propto& 
  \left(1 + \frac14 r^2 \big(3\sin(2\xi)-1 \big)\right) - \frac14 r^2\left(\frac{\pi}{4}\right)^2\big( 1 + \sin(2\xi)\big) \cos(\phi-\bar\phi),
\label{eq:appAzCorrOnly0}
\ee 
where $\zeta_P = (1/2,\,-1)$ for $P = (\pm1,\,0)$, respectively.  It is straightforward to see why the different
polarizations have the same $(\phi - \bar\phi)$ dependence.  The $(\phi - \bar\phi)$ modulation arises from mass-suppressed ``wrong-helicity'' 
($\uparrow\downarrow$ or $\downarrow\uparrow$) $t\bar t$ production.  Once $\Theta$ is integrated out, this spinless
configuration of tops cannot display any residual dependence on the resonance polarization.  

Our discussion thus far has focused on the resonance-squared contribution to the total rate, 
but there is also interference with SM \ttbar\ production.  This is a small 
effect on-peak for narrow resonances, and even for broad resonances if the 
couplings are large compared to QCD.  Still, in a full analysis, one should account
for such effects, especially as they may be numerically important relative to 
the subleading $r^2$ corrections in {\it e.g.} Eq.~\ref{eq:appAzCorrOnly0}.  Let us consider in detail the case of $q \bar q$ annihilation 
into a spin-1 resonance.  The interference comes out proportional to the light quarks' vector
charges (having averaged over forward and backward directions as well as incoming quark chiralities),  
and consists of two terms.  One is a $\cos(\phi+\bar\phi)$ modulation, which simply
tracks the total cross section interference with the SM, and is proportional to the top's vector coupling.
This passes through a zero on-peak, as the resonance propagator goes 90-degrees out of phase
with the gluon propagator.  Thus, we get a sizable interference effect on-peak only if the light quark
and top quark charges each have relatively large vector components, and if the off-peak
contributions in our resonance mass window are highly imbalanced.
The second interference term is proportional to the top axial charge,
and introduces a $\sin(\phi+\bar\phi)$ contribution to the modulation.  This term is suppressed by $\Gamma/M$,
and would have been averaged out in our analysis of Section~\ref{subsec:measurement}, which only looked at $|\phi+\bar\phi|$.
(Note that a sinusoidal modulation component in $(\phi+\bar\phi)$ does {\it not} signal CP-violation,
as it would in $(\phi-\bar\phi)$.)

\section{Spin-2}
\label{app:spin2}

The spin-2 case shares many features with spin-1, including separate
couplings to left and right chirality fields.  We give the lagrangian
for this case in Eq.~\ref{eq:spin2Lag}.  Once again, we parametrize
the relative strengths of the chiral couplings with an angle $\xi$ (Eq.~\ref{eq:xiDef}).
As mentioned in Section \ref{sec:spinCorr}, $q\bar q$ and $gg$
production give rise to $J_{\rm beam}=\pm1$ and $J_{\rm beam}=\pm2$ spin states,
respectively, and each proceed via dimension-five operators.  We therefore
should not prejudice one over the other in general, though $gg$ production
tends to benefit from higher parton luminosities in the case of lighter spin-2 resonances
(less than about 2 TeV~\cite{Frederix:2007gi}),
and $q\bar q$ for heavier resonances.  The 5D angular distributions
are
\be
\frac{d^5\Gamma_{J_{\rm beam}=\pm2}}{d\Omega \, d\bar\Omega \, d\cos\Theta} &\propto&
 \big(\! 1-\cos^4\Theta \big) \Big(\! 1-\cos\theta\cos\bar\theta 
+ \sqrt{1-r^2} \cos(2\xi)  \lp \cos\bar\theta - \cos\theta\big) \Big) \nonumber \\
&&- \sin(2\xi) \sin^4\Theta \sin\theta\sin\bar\theta\cos(\phi+\bar\phi) \nonumber \\
&+& r \Big[ \cos\Theta \sin^3\Theta \big( 1 + \sin(2\xi) \big) \big( \cos\theta\sin\bar\theta\cos\bar\phi + \cos\bar\theta\sin\theta\cos\phi \big) \nonumber \\
&& \hspace{0.7cm} - \sqrt{1-r^2} \cos(2\xi) \cos\Theta \sin^3\Theta \big( \sin\bar\theta\cos\bar\phi - \sin\theta\cos\phi \big) \Big]  \nonumber \\
&-& r^2 \sin^2\Theta \Big[ \cos^2\Theta - \sin(2\xi) - \big( 1 - \sin(2\xi) \cos^2\Theta \big) \cos\theta\cos\bar\theta \nonumber \\
&& \hspace{2cm} + \sin^2\Theta \sin\theta\sin\bar\theta \big( \cos\phi\cos\bar\phi + \sin(2\xi)\sin\phi\sin\bar\phi \big) \Big]
\ee
and
\be
\frac{d^5\Gamma_{J_{\rm beam}=\pm1}}{d\Omega \, d\bar\Omega \, d\cos\Theta} &\propto&
\big( 1-3\cos^2\Theta + 4\cos^4\Theta \big) \Big( 1-\cos\theta\cos\bar\theta \nonumber \\
&& \hspace{4.7cm} + \sqrt{1-r^2} \cos(2\xi) \lp \cos\bar\theta-\cos\theta \big) \Big) \nonumber \\
&&+ \lp 1 -5 \cos^2\Theta + 4 \cos^4\Theta \big) \sin(2\xi) \sin\theta\sin\bar\theta\cos(\phi+\bar\phi) \nonumber \\
&+& \frac12\, r\,\sin(4\Theta) \Big[ \big(1+\sin(2\xi)\big)  \lp \cos\theta\sin\bar\theta\cos\bar\phi + \cos\bar\theta\sin\theta\cos\phi  \big) \nonumber \\
&& \hspace{2.5cm} + \sqrt{1-r^2} \cos(2\xi) 
   \lp \sin\theta\cos\phi - \sin\bar\theta\cos\bar\phi \big) \Big] \nonumber \\
&-& \frac14 r^2 \Big[ 1 -3 \sin (2 \xi ) +2 \cos (4 \Theta)+ \big(1-\sin (2 \xi )\big) \cos (2 \Theta) \nonumber \\
&& \hspace{1cm} - 2 \cos\theta \cos\bar\theta \left[ 1+\cos ^2\Theta-\left(1-7 \cos ^2\Theta+8 \cos ^4\Theta\right)
   \sin (2 \xi )  \right] \nonumber \\
&& \hspace{1cm} + 2 \sin ^2\Theta \sin\theta \sin\bar\theta \Big(  
        \big[1 - \sin (2 \xi ) \left( 1 - 8 \cos^2\Theta \right) \big]  \sin\phi \sin\bar\phi \nonumber \\
&& \hspace{1cm} - \big[1 - 8 \cos^2\Theta - \sin(2\xi) \big] \cos\phi \cos\bar\phi \Big) \Big].
\ee
Just as with spin-1, the $\cos\theta\cos\bar\theta$ distribution lets us distinguish
in principle the difference between vector-like and axial-like couplings.  For all spin-2 polarizations 
the relevant formulas are 
\be
\frac{d\Gamma_V}{d\cos\theta \, d\cos\bar\theta} &\propto& \left( 1 + \frac{2r^2}{3}\right) 
- \left(1 - \frac{2r^2}{3} \right) \cos\theta\cos\bar\theta \nonumber \\
\frac{d\Gamma_A}{d\cos\theta \, d\cos\bar\theta} &\propto& (1-r^2) \big( 1 - \cos\theta\cos\bar\theta \big),
\ee
where $V$ and $A$ are respectively the cases of left-right symmetric
and antisymmetric couplings. We see just a slight modification in the
vector term compared to spin-1, but once again having $M \gtrsim
1$~TeV requires us to fit the distribution to within a few percent
to discriminate between the two cases.

To avoid needing such precision and a full 2D parameter fit, we can integrate the lepton
polar and production angles to find the distributions in $\phi\,,\bar{\phi}$.  The leading term in
$r$ only depends on $\phi + \bar{\phi}$.  We get 
\be
\frac{d\Gamma_{J_{\rm beam}=P}}{d (\phi\, + \bar{\phi})} &\propto&
\Big( 1 - \frac16 r^2 \big( 1 - 5\sin(2\xi)  \big) \Big) + \zeta_P \left(\frac\pi4  \right)^2 \Big( \sin(2\xi) + \frac12 r^2 \big( 1 - \sin(2\xi)  \big)  \Big)\cos(\phi + \bar\phi) \nonumber \\
\frac{d\Gamma_{J_{\rm beam}=P}}{d (\phi\, - \bar{\phi})} &\propto&
\Big( 1 - \frac16 r^2 \big( 1 - 5\sin(2\xi)  \big) \Big) - \frac13 r^2 \left(\frac\pi4 \right)^2 \big( 1 + \sin(2\xi)  \big) \cos(\phi - \bar\phi),
\label{eq:appAzCorrOnlySp2}
\ee
where $\zeta_P = (-2/3,\,1/6,\,1)$ for $P = (\pm2,\,\pm1,\,0)$, respectively.  Once again, we see that the $(\phi - \bar\phi)$-dependence is independent of resonance polarization ({\it cf.} discussion under Eq.~\ref{eq:appAzCorrOnly0}).

The production of a spin-2 particle with $J_{\rm beam}=0$ requires an
operator with dimension greater than five, {\it i.e.}
beyond that of the other polarizations.  It is for this reason that we
did not list any equations for this state in Section \ref{sec:spinCorr}.  However,
we give here the differential decay rate with respect to 
leptonic and production angles, from which one can obtain the 0-polarization portion of 
Eq.~\ref{eq:appAzCorrOnlySp2},
\be
\frac{d^5\Gamma_{J_{\rm beam}=0}}{d \Omega\, d \bar\Omega\, d \cos\Theta} &\propto&
\sin ^2(2 \Theta) \Big( 1-\cos \theta \cos \bar\theta+ \sin (2 \xi ) \sin \theta \sin
   \bar\theta \cos (\phi+\bar\phi) \nonumber \\
&& \hspace{2cm} + \sqrt{1-r^2} \cos (2 \xi ) \sin ^2(2 \Theta) \lp \cos\bar\theta-\cos \theta\big) \Big) \nonumber \\
&-& \frac{1}{3}\, r \Big[ \sin (2 \Theta) \big( 1+3 \cos (2 \Theta)\big) \big( 1+\sin (2 \xi )\big) 
\lb \cos \theta \sin \bar\theta \cos \bar\phi +\cos
   \bar\theta \sin \theta \cos \phi \big] \nonumber \\
&& \hspace{0.6cm} - \frac{1}{2}\, \sqrt{1-r^2} \cos (2 \xi ) \big( 2 \sin (2 \Theta)+3 \sin (4 \Theta) \big) 
\lp \sin\bar\theta \cos \bar\phi -\sin \theta \cos \phi \big) \Big] \nonumber \\
&-& r^2 \Big[ \sin ^2 2\Theta \big( 1 - \sin\theta \sin\bar\theta \cos\phi \cos\bar\phi 
+ \sin(2\xi) (\cos \theta \cos\bar\theta - \sin\theta \sin\bar\theta \sin\phi \sin\bar\phi ) \big) 
\nonumber \\
&& \hspace{0.6cm} -\frac29 \big( 1 + \sin (2 \xi ) \big) \big( 1 + 3\cos^2\Theta \big)
\big( 1 + \cos \theta \cos\bar\theta - \sin\theta \sin\bar\theta \cos(\phi - \bar\phi)  \big) \Big].
\ee


\bibliography{lit}
\bibliographystyle{apsper}

\end{document}